\documentclass[fleqn,usenatbib]{mnras}

\usepackage{newtxtext,newtxmath}

\usepackage[T1]{fontenc}

\DeclareRobustCommand{\VAN}[3]{#2}
\let\VANthebibliography\thebibliography
\def\thebibliography{\DeclareRobustCommand{\VAN}[3]{##3}\VANthebibliography}

\usepackage{graphicx}	
\usepackage{amsmath}	
\defcitealias{Eltvedt_1}{Paper~I}
\defcitealias{Eltvedt_2}{Paper~II}
\defcitealias{Eltvedt_3}{Paper~III}

\title[ACT CMB Lensing]{The VST ATLAS Survey IV: Galaxy, LRG and QSO bias  and HODs via ACT CMB Lensing}

\author[Harshnoor Kaur et al.]{
Harshnoor Kau$r^1$,\thanks{E-mail: harshnoor.kaur@durham.ac.uk (HK)}
T.Shanks$^1$,\thanks{E-mail: tom.shanks@durham.ac.uk (TS)}
S. Bose$^2$
and A.C. Edge$^1$
\\
$^{1}$ CEA, Department of Physics, Durham University, South Road, Durham DH1 3LE, UK\\
$^{2}$ ICC, Department of Physics, Durham University, South Road, Durham DH1 3LE, UK\\
}

\date{Accepted XXX. Received YYY; in original form ZZZ}

\pubyear{2026}

\begin{document}
\label{firstpage}
\pagerange{\pageref{firstpage}--\pageref{lastpage}}
\maketitle

\begin{abstract}
We present angular correlation and power spectral analyses of the clustering of VST ATLAS galaxies, LRGs and QSOs and their cross-correlations with the ACT DR6 CMB lensing map with its $\approx3\times$ higher resolution than Planck DR3. We first  use  $\Lambda$CDM Halofit models at linear scales to estimate galaxy, LRG and QSO bias, $b$,  from the ratio of the 3-D power spectra, $b= P_{gg}/P_{gK}$ and hence the mass clustering amplitudes, $\sigma_8$. For galaxies we find $b_G(z=0.15)=1.22\pm0.07$,  $\sigma_8(z=0)=0.71\pm0.06$ and for LRGs $b_{LRG}(z=0.26)=2.76\pm0.11$,  $\sigma_8(z=0)=0.63\pm0.15$, both consistent with $\Lambda$CDM  predictions. But for QSOs, we estimate a bias of $b_Q(z=1.7)=4.44\pm0.24$ and $\sigma_8(z=0)=0.49\pm0.06$, the latter significantly lower than expected for $\Lambda$CDM and leads to an observed $0.15<z<1.7$ gravitational growth rate faster than predicted for $\Lambda$CDM.  At smaller non-linear scales, HOD model fits imply a broad halo mass distribution for $z\approx0.15$ galaxies but much narrower mass distributions for  LRGs and  QSOs. More surprisingly, the HOD fit for QSOs implies $M_{eff}=5\times10^{14}M_\odot$, $\approx10\times$ higher than expected. Finally, while the LRG and QSO  HOD models  fit  $w_{gg}$ and $w_{gK}$ consistently, the same is not true for the $z\approx0.15$ galaxies unless we hypothesise anti-bias with $b_G\approx0.55$. This result supports previous QSO lensing analyses that also found anti-bias at $r<5$h$^{-1}$Mpc. We suggest that while HOD + NFW halo models may well describe LRG mass profiles and clustering, at intermediate scales the dominant galaxy population (and QSOs) may follow the alternative hierarchical clustering  model of \cite{Peebles1974} where galaxies trace the mass more closely than in a halo model. 
 
\end{abstract}

\begin{keywords}
keyword1 -- keyword2 -- keyword3
\end{keywords}



\section{Introduction}

In \cite{Eltvedt_1,Eltvedt_2, Eltvedt_3} (hereafter \citetalias{Eltvedt_1, Eltvedt_2, Eltvedt_3}) of this series we probed the halo occupation distributions (HODs) of galaxies and QSOs from the VST ATLAS survey using  gravitational lensing data. The first aim had been to use magnification bias of the QSOs themselves to determine the HODs of the foreground galaxies. But we found that by substituting the Planck CMB lensing map  for the background QSOs, that this significantly increased S/N on the galaxy-mass cross-correlation function. In addition, the CMB lensing map allowed us to estimate HODs for the QSOs too, at  a reasonable S/N. However, the Planck lensing map had $\approx6'$ resolution, too low to allow estimates of galaxy or QSO mass profiles. Here we exploit the $3\times$ higher resolution of the ACT lensing map to address both the mass profiles and HODs of galaxies and QSOs.

Our motivation for this work extends back to the discrepancy between the QSO lensing analyses of \cite{BFS1988,Boyle2002,Croom1999,Myers2003,Myers2005}
who found a higher QSO lensing signal for low redshift galaxies than predicted by $\Lambda$CDM and the SDSS results of \cite{Scranton2005,Menard2010}  who found consistency with $\Lambda$CDM. Meanwhile, \cite{Guimaraes2005} confirmed that the lensing convergence associated with galaxy clustering in $\Lambda$CDM N-body simulations was $\approx5\times$ less than observed in the 2QZ QSO lensing studies, in agreement with the \cite{WI98} analysis. \cite{MS2007} further investigated this issue and found that the observed $w_{gK}$ results were consistent with each other, implying that the discrepancy was due  to the use of a HOD model by \cite{Scranton2005} rather than other authors' use of the simpler \cite{WI98} model  which assumes galaxies trace the mass (as did the N-body analysis of \cite{Guimaraes2005}). Here, we shall return to this discussion in the light of the higher S/N of the ACT CMB lensing results and check whether this conclusion still holds.

In terms of models, we further recall that in \citetalias{Eltvedt_1,Eltvedt_2} our galaxy and QSO correlation function predictions were based on the CHOMP halo model lensing package by {\color{blue} C.B. Morrison, M.D. Schneider \& R. Scranton (2013,} \url{https://github.com/morriscb/chomp}). In \citetalias{Eltvedt_3} we reported a problem in modelling QSO HODs with CHOMP in that the $w_{qK}$ predictions were an order of magnitude lower than expected. We shall report below on what causes this problem after comparisons with other packages such as the Core Cosmology Library (CCL, \citealt{Chisari2019}). We note that this issue only affected QSO HOD models and not those of galaxies or LRGs at low redshift.

Finally, as well as the mass profile and HOD results, we shall also estimate the amplitude of the mass power spectrum, characterised by $\sigma_8$, the rms density within a randomly chosen sphere of radius $r<8$h$^{-1}$ Mpc, by first estimating the galaxy/LRG/QSO bias via the ratio of auto- and cross-correlation power spectra at large (linear) scales. Here we shall also consider the angular power spectrum as well as the angular correlation function to help avoid small-scale non-linear contributions at $l>150$ and lens map-making artefacts  at $l<40$ (ACT) \cite{act_lensing_2024} and $l<100$ (Planck, \cite{Planck2018}). We determine the evolution of the mass power spectrum amplitude, $\sigma_8(z)$ in the range $0.15<z<1.7$ and compare our results with those of \cite{Qu_F_2025}.

So in Section \ref{sec:data} we shall describe our VST ATLAS galaxy, LRG and QSO samples along with the ACT and Planck lensing maps. In Section \ref{sec:hods_method},we shall describe the theoretical aspects of our $2\times2pt$ methodology which at large  scales determines bias and $\sigma_8$ and at small and intermediate scales determines the galaxy/LRG/QSO halo mass profiles and HOD parameters. We shall also  describe the lensing model of \cite{WI98} where galaxies are assumed to trace the mass.  In Section \ref{sec:techniques} we describe our observational analysis techniques for both galaxy and QSO samples using the correlation function and the angular power spectrum. Section \ref{sec:halofit} contains our halofit results for bias and $\sigma_8$ from the linear regime. Section \ref{sec:hod_nlb_wi98} presents measurements of bias and HOD parameter estimates for galaxies, LRGs and  QSOs, now analysing auto- and cross-correlation functions in the non-linear regime and finding strong evidence for anti-bias in low redshift galaxies. A possible trade off between anti-bias and $\Omega_m$ is discussed. In Section \ref{sec:discussion} we discuss these results in terms of whether the dominant galaxy population at low redshift is better modelled by a model where galaxies trace the mass than by a halo model. Finally,  in Section \ref{sec:conclusions} we summarise our conclusions. Unless otherwise stated we assume a zero spatial curvature $\Lambda$CDM Universe with $\Omega_m=0.3$, $h=0.7$ and $\sigma_8=0.8$.

\section{Data}
\label{sec:data}
The datasets used  here for galaxies, LRGs and QSOs are similar to those used in \citetalias{Eltvedt_3}. Relevant information on the sample selection is provided in that paper but to make this work self-contained, we summarise  these details below, updating where necessary.

\begin{figure}
\centering
\includegraphics[width=1.0\linewidth]{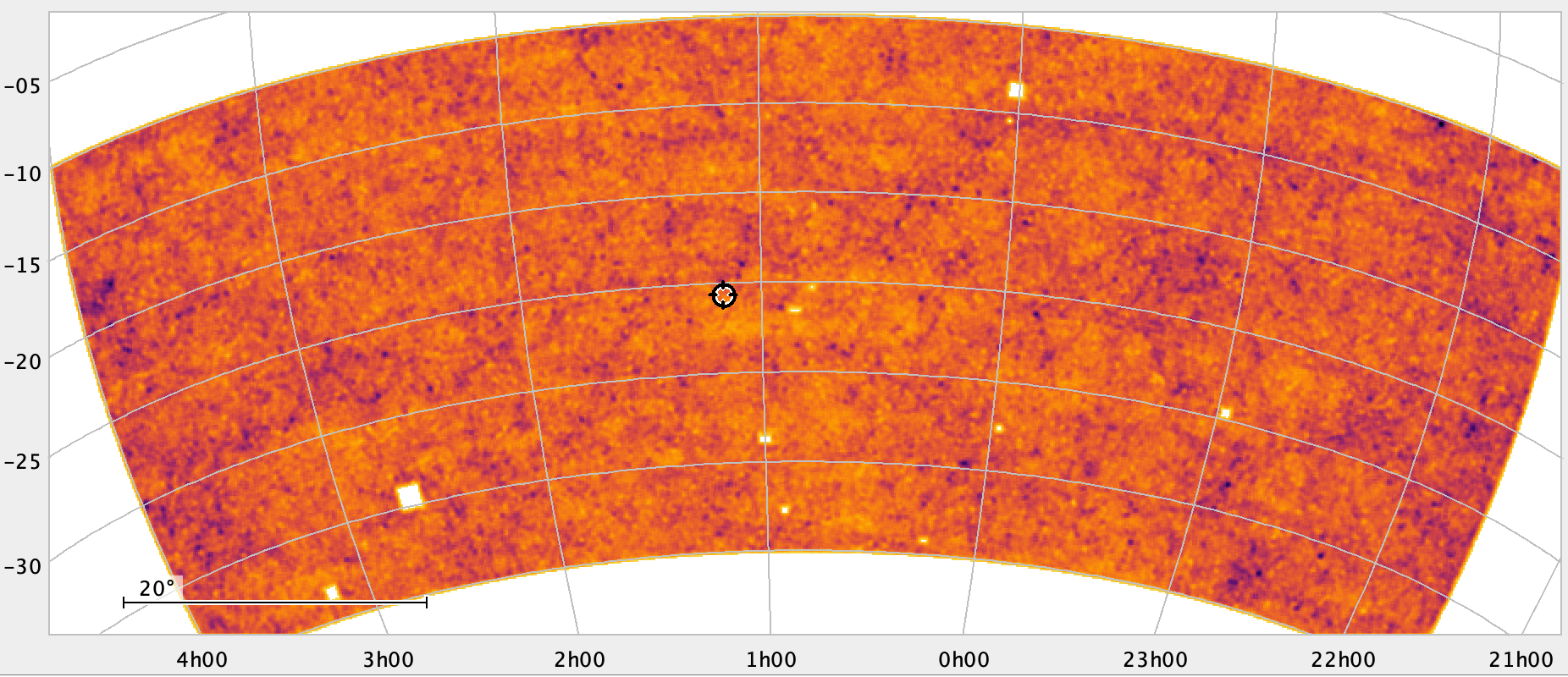}
\includegraphics[width=1.0\linewidth]{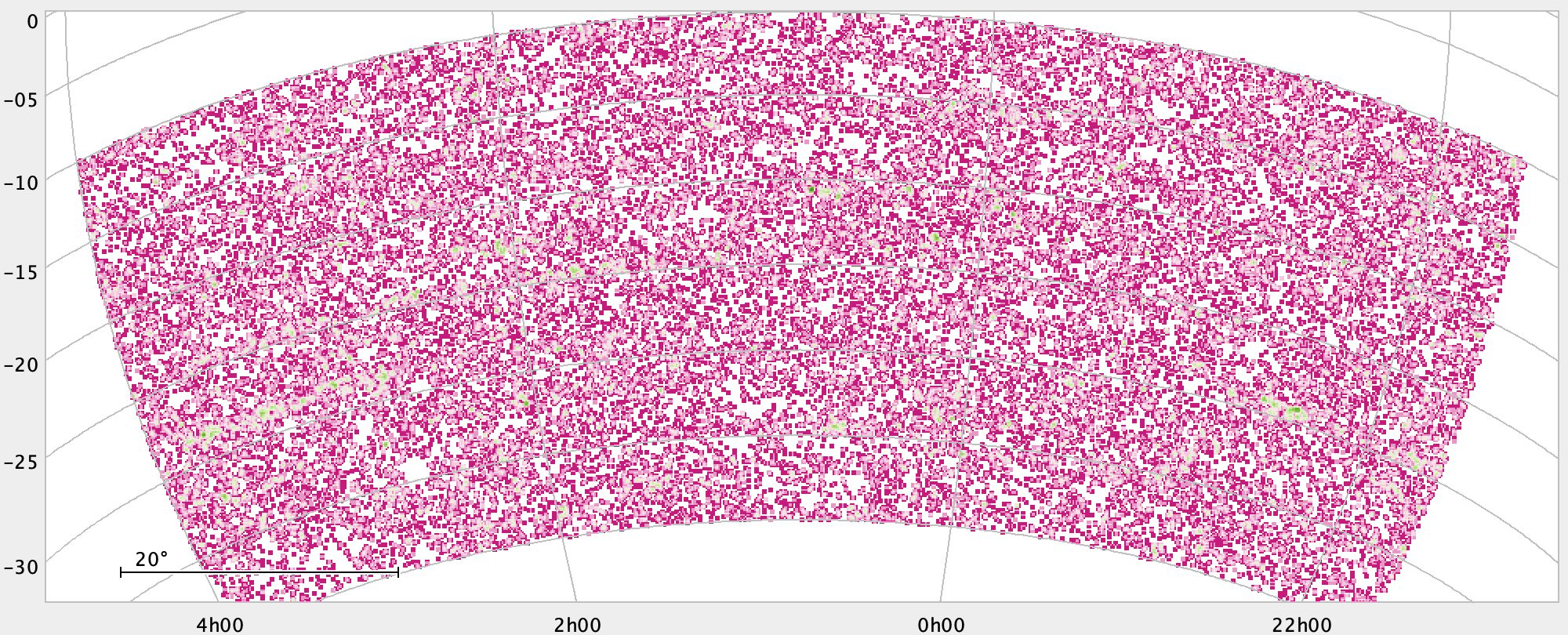}
\includegraphics[width=1.0\linewidth]{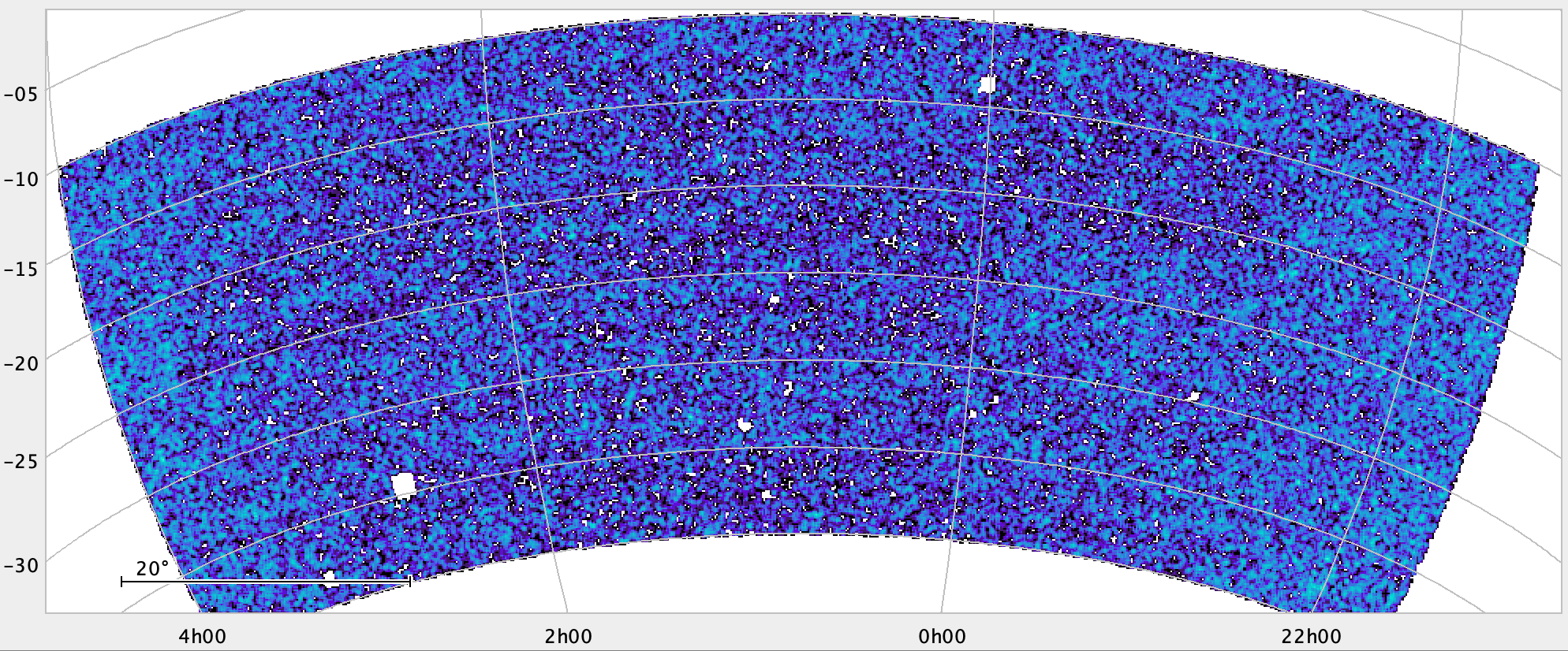}
\caption{(a) Distribution of $17<r<21$ VST ATLAS galaxies in the South Galactic Cap (SGC) in RA, Dec. Darker colours correspond to higher sky density throughout. The sample comprises 6208957 galaxies covering 2700deg$^2$ giving a  galaxy sky density of 2300 deg$^{-2}$. (b) Distribution of $0.16<z<0.36$ VST ATLAS LRGs in the same SGC area, comprising 38511 galaxies, resulting in a sky density of 14.3 deg $^{-2}$.  (c) Distribution of $17<g<22$ VST ATLAS LRGs in the same area. The sample comprises 126052 QSOs giving a  QSO sky density of 46.7deg$^{-2}$. This is lower than the expected  90deg$^{-2}$ at $g<22$ due to extra colour cuts being made to reduce galaxy and star contamination to $\approx7$\%. The 4 SGC subareas are split at $-25<Dec<-10$, $-40<Dec<-25$ and $-37.5<RA<0$, $0<RA<60$. }
\label{fig:quasar_area}
\end{figure}

\subsection{VST ATLAS 17<r<21 Galaxy sample}

The ATLAS galaxy sample has a Gaussian redshift distribution between $0 < z < 0.6$ and  $17 < r < 21$. Since \citetalias{Eltvedt_2} wanted to facilitate comparison with \cite{Scranton2005} who used SDSS data, the galaxies sample has a detection of $r_{sdss} < 21$ using a 0.15 mag offset \cite{shanks2015} to convert to the total r-band SDSS magnitudes, i.e. $r_{sdss} = r_{Kron}-0.15$. Like \citetalias{Eltvedt_3}, we use the improved star-galaxy separation by reclassifying as stars, objects initially classed as galaxies, that lie on the horizontal stellar locus in $r_{kron}-r(<3'')$ vs $r_{kron}$ plot. The same Tycho star and globular cluster masks are used here as in the QSO sample. The galaxy $n(z)$ form in the $17<r<21$ mag range  is well represented by:

\begin{equation} \label{gal_dist}
(\frac{dN}{dz})_G \sim z^{1.3} {\rm exp}[-\frac{z}{0.26}]^{2.17}. 
\end{equation}

\noindent as suggested by \cite{Scranton2005} and confirmed by checking against the number count models of \cite{Metcalfe2006}. The galaxy distribution in the SGC is shown in Fig. \ref{fig:quasar_area}(a).

\subsection{VST ATLAS 0.16<z<0.36 LRG sample}

The Luminous Red Galaxies (LRG) catalogue was created based on the 'Cut 1' selection criteria described in \cite{Eisenstein_2001} for $z \le 0.4$. As in \citetalias{Eltvedt_2} the magnitude limit is  adjusted to  $r<17.9$ to  obtain a sky density coverage of 16deg$^{-2}$ which is close to the 14.3 deg$^{-2}$ sky density coverage obtained by \cite{Eisenstein_2001}.(see Fig. \ref{fig:quasar_area}(b)).  The LRGs have an approximately flat redshift distribution between $0.16 < z < 0.36$. For technical issues with  CCL, which prefers Gaussian to top-hat redshift ranges, we model this redshift distribution with a flat Gaussian,

\begin{equation}
(\frac{dN}{dz})_{LRG} = \exp(-((z-0.26)/\sqrt 2\sigma)^2))/\sqrt(2\pi)\sigma, 
\end{equation}
\noindent with $\sigma=0.3$ and cut in the range $0.16<z<0.36$.

\subsection{VST ATLAS QSO sample}

The QSO sample is selected from the VST ATLAS survey \citep{shanks2015} which covers a total area of $\approx 4700$ deg$^2$ in the Southern Hemisphere ($Dec<0$ deg), split into a  2700deg$^2$ SGC area (see Fig.\ref{fig:quasar_area}) and a 2000deg$^2$ NGC area. As noted above, this work only uses the Southern Hemisphere QSO sample as the ACT survey only overlaps the VST ATLAS survey in its  2700deg$^2$ SGC area.  The QSO sample is defined between $17 < g < 22$ mag and covers photometric redshift range of $1 < z < 2.7$ with peak at $z = 1.7$. The photo-z distribution of the sample (see Fig. 15 of \citetalias{Eltvedt_1}) was obtained by applying  ANNZ2 to the VST ATLAS and  WISE broad-band colours (\citetalias{Eltvedt_3}). The accuracy of this photometric redshift distribution has been checked in comparisons with SDSS and DESI data. The QSO $n(z)$ form is well 
represented by:


\begin{equation} \label{qso_dist}
(\frac{dN}{dz})_Q \sim z^{2.56} {\rm exp}[-\frac{z}{2.02}]^{12.76}. 
\end{equation}

\noindent as suggested by  \cite{Scranton2005}) for their $17<g<21$ QSO sample and confirmed for our $17<g<22$ sample by  \citetalias{Eltvedt_3}.

As described in \citetalias{Eltvedt_3}, Tycho star and globular cluster+bright galaxy  masks are applied to the QSO data. The masked "holes" so produced  can be dealt with exactly in $w_{qq}$ and $w_{q\kappa}$ analyses by use of random catalogues with $>10\times$ the sky density of the QSOs.

The quasar sample has many photometric colour cuts and selection criteria applied to remove contamination. Some of the main cuts applied are described below but more detail can be found in \citetalias{Eltvedt_1,Eltvedt_2,Eltvedt_3}. After a first selection to cut out White Dwarfs is applied, UVX and mid-IR colour cuts are applied, following \citetalias{Eltvedt_1}. Then following \citetalias{Eltvedt_3}, we make a further cut to reduce contamination at the price of also reducing QSO completeness:-

\begin{equation} \label{uvx_cut_2}
    \begin{gathered}
    -0.25 < (g-r) < 0.4\  \\
    (u-g) < 0.55\\ 
    (r-W1) < 5\\
    \end{gathered}
\end{equation}

\noindent These last cuts are performed to reduce the possibility of galaxy contamination, based on an analysis of spectroscopically confirmed QSOs in \citetalias{Eltvedt_3}. They leave the QSO sample with a  sky density of $\approx47$ deg$^{-2}$ which is closer to the sky density expected at $g<21$ and roughly half the sky density in the  $g<22$ mag limited initial sample.

Then a further mask to remove Tycho stars with \(V_T < 12.5\), is applied according to \cite{BA_2023}.  Globular clusters, dwarf galaxies and areas with poor photometry are also similarly masked. This corresponds to some of the holes seen in Fig. \ref{fig:quasar_area}(c) which shows the RA and Dec ranges of the QSO sample in the SGC.


\begin{figure}
\centering
\includegraphics[width=1.0\linewidth]{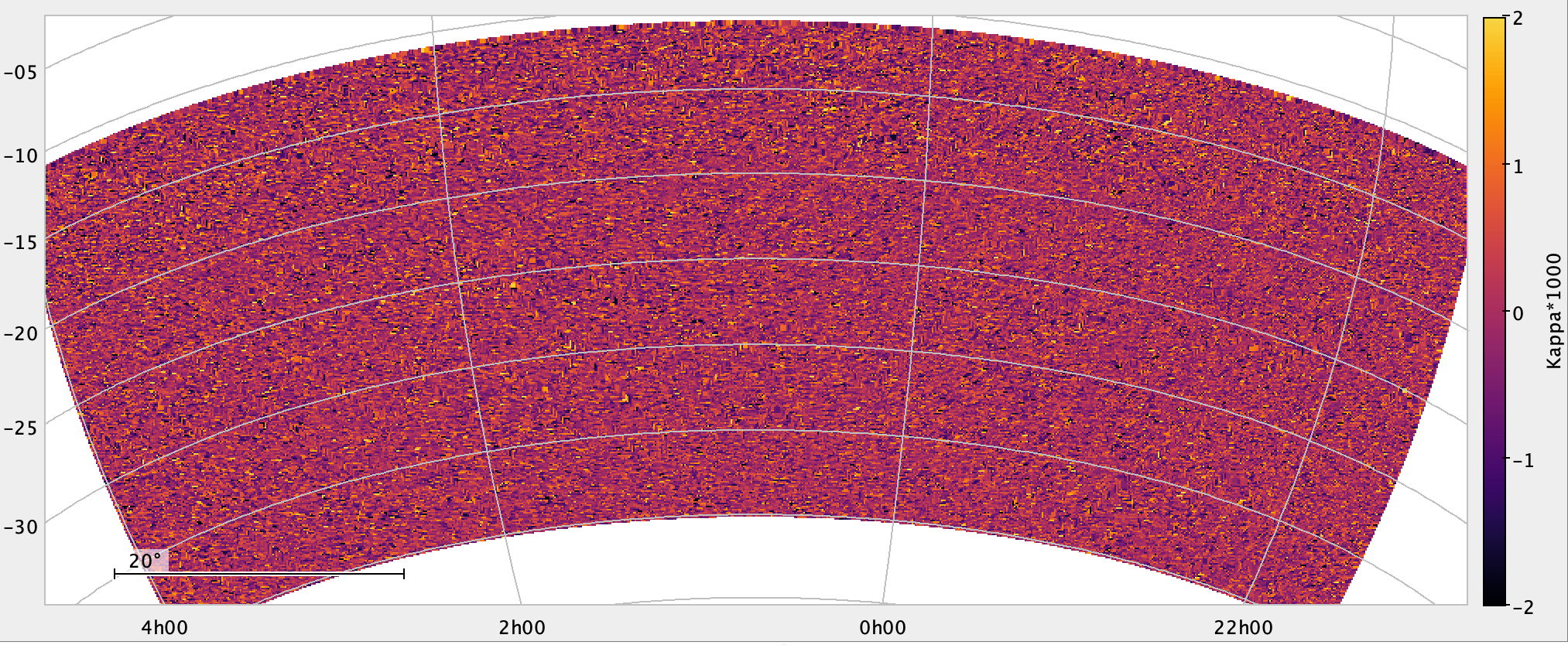}
\caption{ACT DR6 lensing convergence map \protect{\citep{act_lensing_2024}} output using {\it healpy alm2map} with parameters of Healpix \citep{Gorski2005} $nside=4096$ and $lmax=4096$, after pre-filtering   the $a_{lm}$ map with a 2-D Gaussian filter of $FWHM=3'$. Bright areas correspond to peaks of the mass distribution and dark areas correspond to dips.
}
\label{fig:act_area}
\end{figure}

\subsection{Atacama Cosmology Telescope CMB Lensing Map}
\label{sec:ACTmap}

The CMB lensing map is obtained from the ACT DATA Release 6 (DR6,  \citealt{act_lensing_2024}),  which covers an area of 9400 deg$^2$ with 7700deg$^2$ lying at $Dec<0$ deg and of which 2700deg$^2$ overlaps the VST ATLAS area in the Southern Galactic Cap (SGC).  ACT obtains CMB temperature and polarization data  at approximately 97 and 149 GHz using the Advanced ACTPol (ADvACT) instrument. This data had been transformed into a lensing convergence ($\kappa$) map using the pipeline developed for the Simon Observatory \citep{Simons_Obseratory_2019}.

We obtained the lensing baseline map from the ACT DR6 website. This map is stored in spherical harmonics, $a_{lm}$. We use the {\it healpy alm2map} routine  to convert its Healpix \citep{Gorski2005} pixels to RA and Dec equatorial coordinates at $nside = 4096$ and $l_{max} = 4096$ for estimating angular cross-correlation functions, $w_{q\kappa}$. We use the same data to estimate the angular power spectrum but now with $nside=1024$ since we are working at larger scales when using $C_l$'s. The $a_{lm}$ map is also pre-smoothed using a FWHM = 3 arcmin Gaussian beam, a width chosen to optimise the S/N in the cross-correlation functions following the method of \citetalias{Eltvedt_3} for the Planck lensing map. We then apply the mask provided by \cite{act_lensing_2024} and analyse the 2700 deg$^2$ area that overlaps our VST ATLAS SGC Galaxy, Luminous Red Galaxy and QSO samples (see Figs. \ref{fig:quasar_area}, \ref{fig:act_area}).

As noted by \cite{act_lensing_2024}, the ACT lensing map is pre-filtered by removing all modes outside the range $40<l<3000$ and so this filter is always applied to our angular cross-correlation function models.

\subsection{Planck Lensing Map}
In terms of the Planck CMB lensing map, here we use the same DR3 data as used by \citetalias{Eltvedt_3} with the same steps to produce the convergence map in equal area pixels as described in their Section 2.2. Thus, for the Planck map we use $nside=2048$ and $l_{max}=4096$ and a Gaussian kernel smoothing scale of $FWHM=15'$.  \cite{Carron2018} (via \citealt{Geach_2019}) recommends an  $l>100$ limit for this DR3 Planck map. Therefore, for angular power spectrum ($C_l$) analyses we only used the range $l>100$ for Planck whereas \cite{Qu_F_2025}  use $l>50$. When we consider that we aim to fit only  in the linear regime this means we are restricted to $100<l<150$ for Planck analyses. We also now assume that the Planck map is not pre-filtered at any scale.

\section{Galaxy and mass clustering route to bias, \protect{$\sigma_8$} and HODs}
\label{sec:hods_method}
\subsection{Bias and $\sigma_8$ by combining galaxy auto- and cross-correlation functions}
Linear bias is defined by $\xi_{gg}=\xi_{mm}\times b^2$ or equivalently $P_{gg}(k)=P_{mm}\times b^2$ where $\xi$ and $P$ are the 3-D 2-point correlation function and power spectrum of the distribution of mass ($m$) and $b$ is assumed to be constant at fixed $z$ in the linear regime. Finding $b$ by comparing the amplitude of galaxy or QSO clustering to the amplitude of the mass power spectrum has been a basic route to galaxy and quasar halo masses via clustering. However, originally, the amplitude of the mass power spectrum  had to be assumed, inferred from the amplitude of CMB temperature  fluctuations at $z\approx1100$. With the availability of CMB lensing convergence ($\kappa$)  maps we can now estimate the bias by measuring $P_{gg}=b^2P_{mm}$ and $P_{gm}=b\times P_{mm}$ and estimating $b$ by forming $P_{gg}/P_{gm}=b$. Previously we have followed this method to estimate galaxy bias  by substituting QSO lensing maps for the CMB map while others have substituted weak shear lensing maps of background galaxies but the CMB route seems to have more statistical power than the QSOs and fewer systematics than the shear maps motivating our use of Planck and ACT lensing maps here.

To measure $P_{gg}$ and $P_{gm}$ we shall follow \cite{Qu_F_2025} and initially use the angular power spectra $C_l^{gg}$ and $C_l^{g\kappa}$ since we wish to focus on the linear regime to make these measurements and secondly CMB lensing maps have artefacts due to pre-filtering that are more easily dealt with in $l$-space. However, masks in eg QSO surveys can be more difficult to deal with using $C_l$'s and so we shall also use $w_{gg}$ and $w_{g\kappa}$ to cross check the amplitude of our auto- and cross-correlations. So we will be calculating the angular auto- and cross-correlation functions for the galaxies, LRGs and QSOs. Here we shall take into account in particular the $40<l<3000$ top-hat pre-filter of the ACT lensing map (see Sect. \ref{sec:ACTmap} above) by applying a similar filter to our models for $w_{g\kappa}$  We shall therefore start by fitting a bias, $b_A$ to the autocorrelation function and another bias, $b_X$ to the cross-correlation function by assuming the same unbiased Halofit power spectrum for both. Since $P_{gg}=b_A^2P_{mm}$ and $P_{gm}=b_X\times P_{mm}$, then $P_{gg}/P_{gm}=b_A^2/b_X=b$ provides our estimate of the galaxy bias, $b$. Since $P_{gg}=b^2\sigma_8^2P_{mm}$ then to maintain the fit to $P_{gg}$, $b*\sigma_8'(z) = b_A*\sigma_8^{fid}(z)$ so $\sigma_8'(z)=\sigma_8^{fid}(z)*b_A/b$. Therefore $\sigma_8'(z)=\sigma_8^{fid}(z)*b_X/b_A$. Usually we shall assume a fiducial $\sigma_8^{fid}(z=0)=0.8$ and calculate the new $\sigma_8'(z)=0.8*D(z)*b_X/b_A$ where $D(z)$ is the linear mass growth factor between $z$ and $z=0$, assuming $\Lambda$CDM (where $D(z)=1/(1+z)$ for the Einstein-de Sitter model). Thus we obtain $\sigma_8'(z)$ as well as $b(z)$ from our auto- and cross-correlation function analysis.

\subsection{Mass profiles and HOD's via auto- and cross-correlation functions}
We shall also be exploiting the increased resolution of the ACT lensing map over Planck's to fit the 1-halo term as well as the 2-halo term in order to make improved fits of the galaxy and QSO Halo Occupation Distributions. Here we shall mainly follow \citetalias{Eltvedt_2} in their cross-correlation of the Planck map with VST ATLAS galaxy and QSO surveys. However, we shall be using CCL instead of CHOMP to produce our HOD models since we now find CHOMP is less accurate in dealing with higher redshift surveys such as QSOs. We shall also be taking into account the effect of the ACT beam as well as the $40<l<3000$ filter in our ACT cross-correlation analyses. We shall also be finding best  MCMC fits of the HOD parameters rather than simply giving examples of typical HOD model fits as in \citetalias{Eltvedt_2}. We shall therefore be using the \cite{Z2007} HOD parameterisation and fitting the 5 parameters $M_{min}$, $M_0$, $M_1'$, $\alpha$ and $\sigma_M$. This will lead to a galaxy/LRG/QSO halo mass distribution and a new effective halo mass for galaxies at $z=0.15$, LRGs at $z=0.26$ and  QSOs at $z=1.7$ for comparison with the masses from  the 2-halo bias estimates from the Halofit models discussed above.

Clearly the cross-correlation of the CMB lensing mass map with low redshift galaxies, $w_{g\kappa}$ is essentially a stacked mass profile of the galaxies since at $z=0.15$, the $2-3'$ resolution of the ACT map corresponds to $\approx0.2-0.3$h$^{-1}$ Mpc, probing the 1-halo term at $<1$h$^{-1}$ Mpc scales. So as well as fitting HOD parameters we shall also be able to check the accuracy of fit of our assumed NFW profile for the 1-halo term in the \cite{Z2007} model. Even for $z=1.7$ QSOs the $2-3'$ ACT resolution corresponds to $\approx0.7-1$h$^{-1}$Mpc and so even without detailed  modelling of the ACT PSF, some further testing of the NFW assumption for the 1-halo term may be possible even here.

\subsection{Lensing model where galaxies trace the mass}
Here we describe the lensing model of  \cite{WI98} where galaxies are assumed to trace the mass. In this model, the galaxy auto-correlation function will have the same form as the that of the mass, down to the galaxy  linear bias  $b$,  implying $w_{gg}=b^2*w_{mm}$. Then, in the case of foreground galaxies lensing background QSOs, it can be shown that:-
$$w_{gq}=(2.5\alpha-1)*2\kappa/b*w_{gg}$$
\noindent where lensing convergence $\kappa$ is given by:-
$$\kappa=\rho_{crit}\Omega_m\int_{z_{min}}^{z_{max}}(cdt/dz)*(1+z)^3/\Sigma_{crit}dz$$ (=0.016 for $z=1.7$ QSO lens  sources) and $\alpha$ is the slope of the QSO number counts in the relevant magnitude range. Note that these WI98 lensing equations are the same as those used by CCL except that WI98 assume the observed form of $w_{gg}$ whereas CCL derive it from an assumed linear $P(k)$ (or Halofit) and then project these using Limber's formula.

Now when \cite{Myers2003,Myers2005} compared this prediction for $w_{gq}$ to data where QSOs were the lensing sources, they found a significantly lower result than was subsequently found for the basic HOD model of \cite{Scranton2005}.
For CMB lensing by $17<r<21$ mag galaxies, we calculate $\kappa=0.026$ for $0.1<z<0.36$ which implies $w_{g\kappa}=\kappa/b=0.026/b*w_{gg}$. Thus we estimate $b$ by comparing this relation with the observed $w_{g\kappa}$ by moving $w_{gg}$ down by this $\kappa/b$ factor as shown e.g. in Fig. \ref{fig:wgg-wgk-hod} and discussed in more detail in Sect. \ref{sec:discussion}. For $0.16<z<0.36$ LRGs we calculate $\kappa=0.026$, again allowing the bias to be calculated from $w_{g\kappa}=\kappa/b=0.026/b*w_{gg}$.

\section{\protect{$2\times2pt$} Correlation analyses of galaxies, QSOs and CMB lensing maps - techniques}
\label{sec:techniques}

\subsection{Angular correlation functions and power spectra}
We use both the 2-D angular correlation function $w(\theta)$ and the 2-D angular power spectrum, $C_l$, in these analyses. Although these are related by a spherical harmonic transform, they have advantages and disadvantages in terms of their estimation properties. The $C_l$ modes are statistically independent for whole sky coverage whereas this is not so for the angular correlation function. However, for masked data with less than whole sky coverage the $C_l$ modes show increasing covariance and also require normalising by the fraction of sky covered, $f_{sky}$. More complicated masks where bright stars and galaxies have also been masked and these resulting  smaller "holes" can only be addressed using realistic simulations. The raw modes before normalising, are called pseudo-$C_l$'s denoted by ${\widetilde{C}}_l$. Here we shall use the approximate result ${\widetilde{C}}_l = N\times f_{sky}\times w_2$ where $w_2=\int_{4\pi} W(\hat{{\bf n}})^2 d{\bf n}$ and $f_{sky}$ is the fraction of sky covered and $W({\bf n})$ is the window function (see eqs 9, A32 of \cite{Hivon2002}). This relation  is exact if the underlying  spectrum is the white noise spectrum, $<C_l>=N$.  In this case, $C_l={\widetilde{C}}_l*4\pi/w_2={\widetilde{C}}_l*4\pi/\int_{4\pi} W(\hat{{\bf n}})^2 d{\bf n}$. In the cross-correlation case this becomes $C_l={\widetilde{C}}_l*4\pi/\int_{4\pi} W_\kappa(\hat{{\bf n}})W_g(\hat{{\bf n}}) d{\bf n}$ and this is the relation we apply to convert pseudo-$C_l$ into $C_l$'s. However, in this work we generally used the same "boundary defining" mask in each case. Since our spectra are not white noise, these normalisations are only approximate. So, rather than testing via simulations, here we simply check whether the $C_l$ results are consistent with those from $w(\theta)$. We generally find excellent agreement, indicating that this approximation works for the angular power spectra considered here. 

The maps of $\delta\rho/\bar{\rho}$ for galaxy/QSO pixels and $\delta\kappa/\bar{\kappa}$ for the lensing pixels are then created. We note that the mask supplied by ACT for the lensing map is then applied before  the mask for the ATLAS SGC sub-area perimeter. With $NSIDE=1024$, we find that the average $\kappa$ per pixel for ACT is $\bar{\kappa}=0.00016$ and $\bar{\rho_g}=17.0$ and $\bar{\rho_q}=0.16$ in the SGC. For Planck the average $\kappa$ per pixel  is $\bar{\kappa}=4.075\times10^{-4}$ in the NGC and $\bar{\kappa}=4.075\times10^{-4}$ in the SGC with $\bar{\rho_g}=17.0$ and $\bar{\rho_q}=0.16$ in the NGC. Then the healpy apodizing function  is applied with a 5 degree scale length and finally  the pseudo-Cl's are estimated using the healpy anafast routine. To plot $C_l$ results in log bins we simply average the individually estimated $C_l$ modes within bins equally spaced in  $\log_{10}l$. We then first  estimate $C_l^{gg}$ (or similarly $C_l^{qq}$) by  applying healpy anafast to the squared product of each of these masked, apodized, galaxy/LRG/QSO maps.


To estimate $w(\theta)$ we follow \citetalias{Eltvedt_3} and use the Landy-Szalay \citep{LS1993} estimator for the galaxy and QSO auto-correlations while using the healpy routine {\it anafast} for the cross-correlations with the ACT and CMB lensing maps. The random galaxy, LRG and QSO distributions have $>10\times$ the number of actual galaxies/LRGs/QSOs and the bright star and galaxy mask is applied to the randoms in exactly same way it is applied to the data. While for the galaxies it was possible to create both $C_l$'s and $w(\theta)$'s down to sub-arcmin scales, for LRGs and QSOs the $C_l$'s were dominated by the random noise term so we chose to use  only $w_{gg}$ and $w_{qq}$  for our analysis here.

\subsection{Galaxy-CMB cross-correlation}
To estimate $C_{g\kappa}^l$, we proceed by using the list of CMB lensing pixels with $NSIDE=1024$,  as for $C_{gg}^l$ . $C_{g\kappa}^l$ for cross-correlating galaxies and CMB lensing $\kappa$ is then created by applying healpy anafast to the product of the masked, apodized, galaxy/QSO map and the masked, apodized CMB lensing map. The main difference with the auto-correlation power spectra  is that instead of $W(\hat{{\bf n}})^2$ in converting from pseudo-$C^l$'s to $C^l$'s, we use  $W_g(\hat{{\bf n}})\times W_\kappa(\hat{{\bf n}})$ where $W_g$ represents the galaxy or QSO mask and $W_\kappa$ the lensing map mask. But again note that in this work we generally used the same `boundary defining' apodized mask in each case. To plot $C^l$ results in log bins we again simply average the individually estimated $C^l$ modes within bins equally spaced at the same log intervals as above so that the auto- and cross-correlation power spectra are binned consistently. 

We estimate $w_{g\kappa}$ following \citetalias{Eltvedt_3}. We first create the list of CMB lensing pixels with $NSIDE=l_{max}=4096$ in the case of ACT, giving equal area $\kappa$ pixels $\approx 1'$ in extent. For the galaxies/QSOs we do not use the healpix pixels and simply loop through all the galaxies/QSOs and average  all the $\kappa$ pixels at distance $\theta$ from each galaxy/QSO. So the estimator here simplifies to:



$$w_{g\kappa}(\theta)=\Sigma_{i,j}\mathbf{|\hat{n}}_i|\delta_\kappa(\mathbf{\hat{n}}_j)$$. 

\noindent where $\mathbf{\hat{n}_i}$ and $\mathbf{\hat{n}_j}$ represents the positions of  galaxy $i$ and convergence pixel $k$ respectively. With $NSIDE=4096$, we find that the average $\kappa$ per pixel is $\bar{\kappa}=0.00016$. Note that by centring on galaxies and looking for CMB lensing pixels some systematic artefacts like  density gradients in the sky distribution of galaxies/QSOs can be reduced, although such systematics in the CMB map will still affect the results.

In modelling $w_{g\kappa}$ we must apply a top-hat filter in the range $40<l<3000$ to account for the pre-filtering of the ACT lensing map in this way. It does not appear that any similar filter has been applied to the Planck data. In all cases we estimate errors by splitting the SGC and NGC data each into 4 sub-areas and taking the error on the mean of these 4 measurements (see \citetalias{Eltvedt_3}). For Planck we then make a combined estimate by taking an error weighted mean of the NGC and SGC measurements.


\section{Halofit bias estimates for \protect{$17<r<21$} ATLAS galaxies, LRGs and $z\approx1.7$ ATLAS QSOs.}
\label{sec:halofit}

\subsection{Auto- and cross-correlation results for 17<r<21 galaxies}
We show in Fig. \ref{fig:clggclgk} the angular power spectra for $17<r<21$ galaxies' auto- and galaxy-ACT lensing map cross-correlation functions $C_{gg}^l$ and $C_{g\kappa}^l$ in blue and similar results for Planck in red. These  have been renormalised from the pseudo-$C^l$ as described in Section \ref{sec:techniques} above.
The ACT and Planck results look quite consistent when we take into account  the higher resolution of the ACT data causing the ACT $C_{gk}^l$ result to have more power than the Planck equivalent at $l>200$. The $C_{gg}^l$ for ACT only includes the SGC galaxies whereas the Planck data includes N and S and this leads to increased power being seen in ACT at $l>200$, probably due to differences in the masking rather than statistical fluctuations given the small size of the field-field errors. The vertical dashed lines show the range we shall be fitting Halofit models and we note that this range lies at $l<200$ so that these large-scale cosmological results are unaffected by either of these slight differences.

\subsection{Halofit bias estimates for \protect{$17<r<21$} ATLAS galaxies}

\begin{figure}
	\includegraphics[width=\columnwidth]{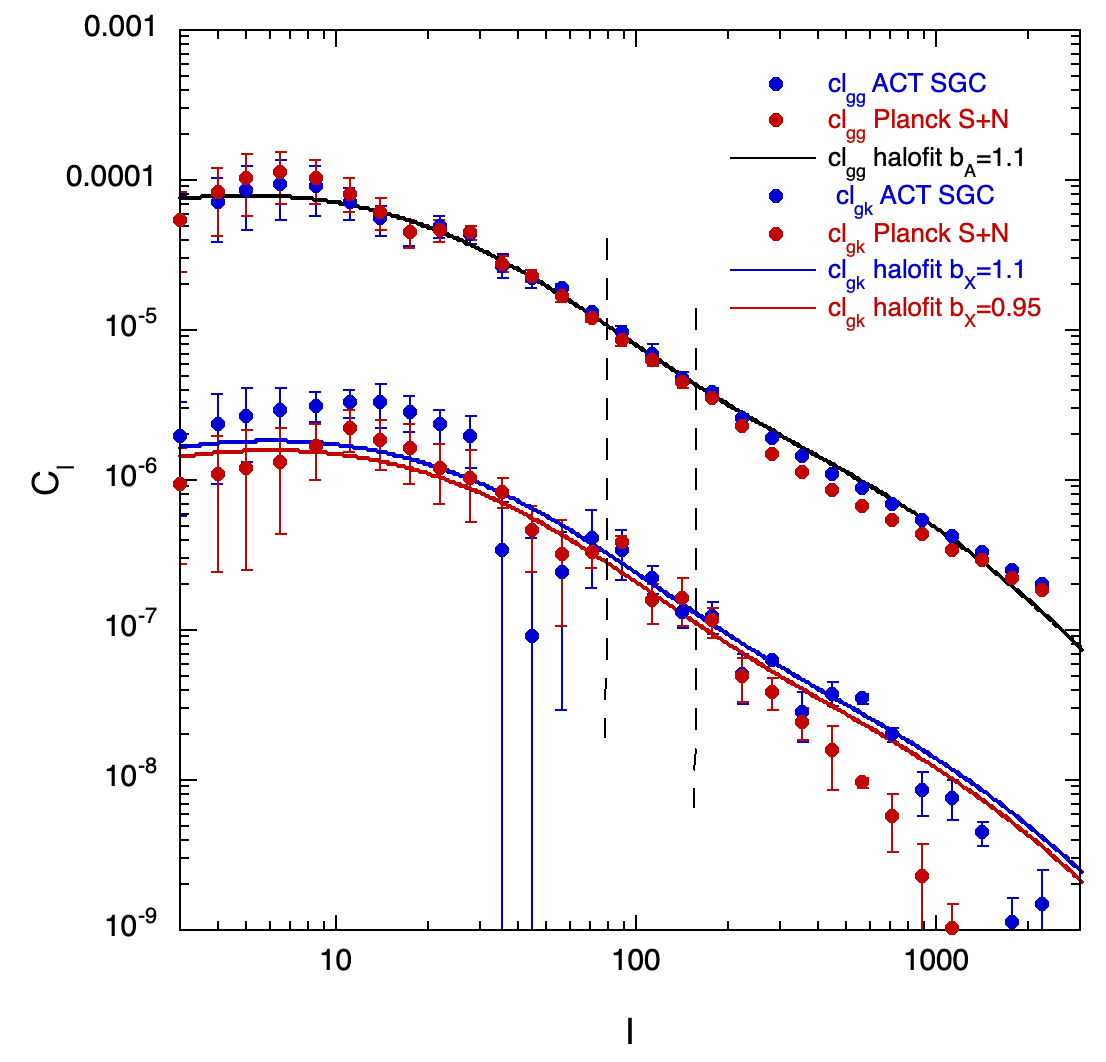}
   \caption{$C_{gg}^l$ and $C_{g\kappa}^l$ models are based on HALOFIT power spectra from CCL and compared to the observed angular auto- and cross- power spectra in the ACT SGC and Planck NGC + SGC lensing map areas. Note that for the ACT fit all 3 points between the dashed lines are fitted ($80<l<150$) whereas in the Planck case only the two with $100<l<150$ are fitted. The fits shown are for $b_A$ and $b_X$ (see Table \ref{tab:clgk}).}
   \label{fig:clggclgk}
\end{figure}

Here we assume a simple Halofit model with constant bias, $b$, to use the result $P_{gg}=b^2P_{mm}$ and $P_{gm}=bP_{mm}$ giving $P_{gg}/P_{gm}=b$ to estimate bias and thus estimate $\sigma_8$. We fit the Halofit model to $C_{g\kappa}^l$ in the ACT range $80<l<150$, i.e. $7<r<13$h$^{-1}$ Mpc. We chose this range because at  $l>150$ we are  getting too close to the 1-halo term and although ACT data is nominally usable to $l>40$, the scales at $40<l<80$ are becoming less reliable in terms of ACT lensing map systematics. Indeed, Planck lensing map data is only reliable at $l>100$ \citep{Carron2018} so the slightly smaller range of $100<l<150$ is only used in this 2-halo term range. The $C_{gg}^l$ are fitted in the same ranges for the ACT and Planck cases in each case so that their ratio will be as consistent as possible. For ACT, the best fit to $C_{gg}^l$ with $\sigma_8(z=0.15)=0.8*0.927=0.74$ is $b_A=1.10$ and in the same  $l$ range the best fit to $C_{gK}^l$ is $b_X=1.10$ again with $\sigma_8=0.74$.  This implies $b=b_A^2/b_X=1.1$ and $\sigma_8'(z=0.15)=\sigma_8(z=0.15)*b_X/b_A$ (where $'$ denotes new value as opposed to original value) and so  here we find  $b=1.1^2/1.1=1.1$ and $\sigma_8'(z=0.15)=0.74$. Both $C_{gg}^l$ and $C_{g\kappa}^l$ are therefore well-fitted with $b(z=0.15)=1.1$ and $\sigma_8(z=0.15)=0.74$ (see Fig. \ref{fig:clggclgk}). The implied value of $\sigma_8'(z=0)=0.74/0.9266=0.8$ is unchanged in this case. In the Planck case, the Halofit model is only fitted to the range $100<l<150$ and here we find a slightly lower value of $b_X=0.95\pm0.15$. Given the same value of $b_A=1.1\pm0.05$ applies this leads to a higher value of $b=1.27\pm0.21$ and thus a lower value of $\sigma_8'(z=0.15)=0.64\pm0.10$. Since all fitted values are within the errors we average these to find $b=1.16\pm0.13$ and  $\sigma_8'(z=0.15)=0.69\pm0.08$. These final values are close to the initial value of $b=1.1$ fitted to ACT in Fig. \ref{fig:clggclgk} and the results are summarised in Table \ref{tab:clgk}.

\begin{figure}
    \includegraphics[width=\columnwidth]{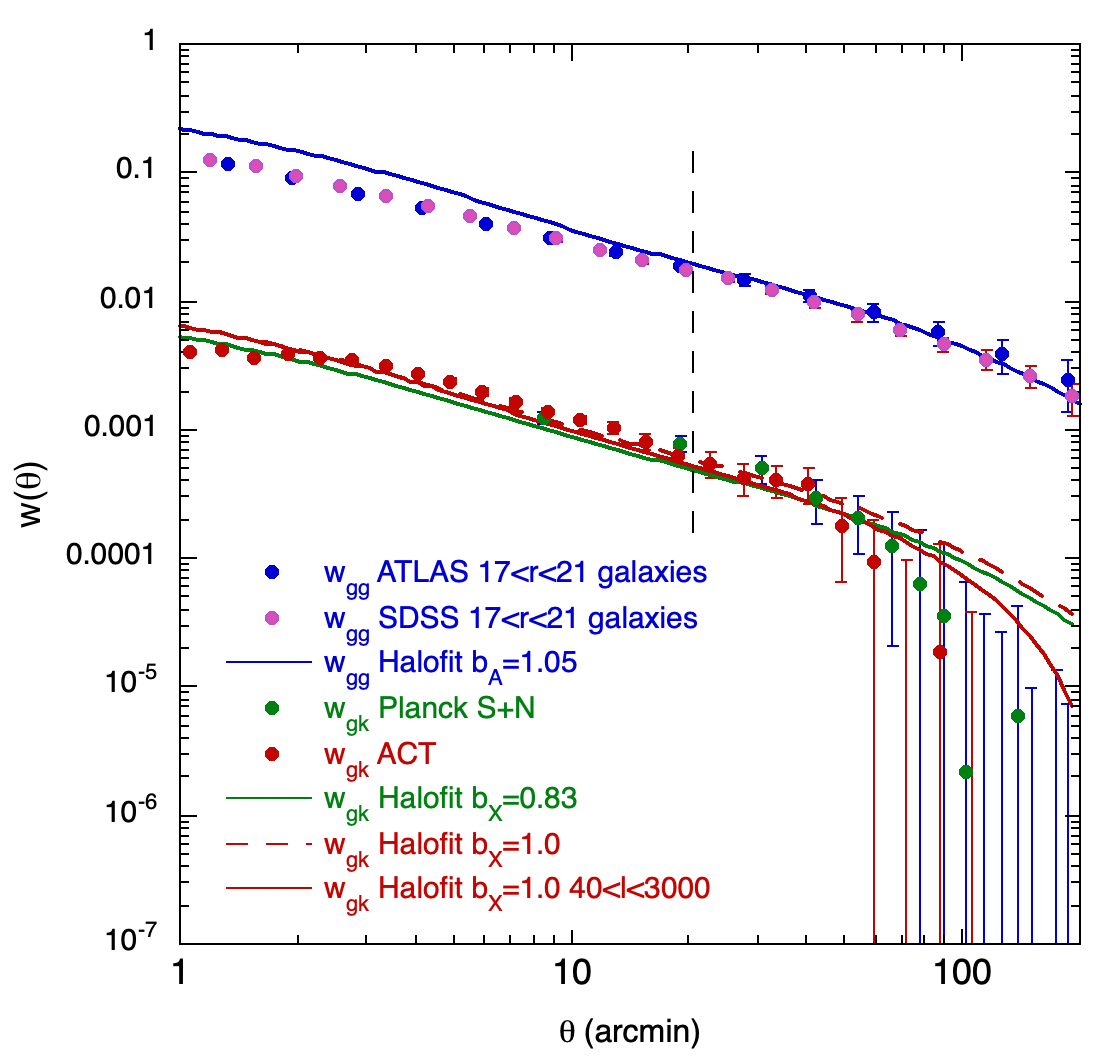}
    \caption{The auto- and cross-corelation angular correlation functions $w_{gg}$ and $w_{g\kappa}$ for $17<r<21$ galaxies. Models are based on HALOFIT power spectra from CCL. Fits are made only to scales larger than indicated by the dashed line, $\theta>20'$. The auto-correlation function, $w_{gg}$, is best fit by bias $b_A=1.05$ assuming $\sigma_8=0.8$. With the same assumption, the Planck $w_{g\kappa}$ is best fit by bias $b_X=0.83$ and  the ACT $w_{g\kappa}$ by $b_X=1.0$ when the model is pre-filtered in the range $40<l<3000$ to match the ACT data.}
    \label{fig:wggwgk}
\end{figure}

Fig. \ref{fig:wggwgk} shows $w_{gg}$ and $w_{g\kappa}$ for the $17<r<21$ galaxy sample and the above parameters of $b(z=0.15)=1.1$ and $\sigma_8(z=0.15)=0.74$ are also shown to give good fits in these cases. Indeed, these parameters are also the best fits to these data when fitting in the range $3'<\theta<100'$ and this consistency supports the approximations we have made particularly in the $C^l$ case.

Taking error-weighted averages over the Average values in Tables \ref{tab:clgk} and \ref{tab:wgk} we obtain our final estimates of  $b_g=1.22\pm0.07$ and $\sigma_8'(z=0.15)=0.66\pm0.053$. We summarise these results in Fig. \ref{fig:sigma8_z}.

\begin{table}
	\centering
	\caption{Fits to $17<r<21$ galaxy $C_{gg}^l$ and $C_{g\kappa}^l$ for bias $b_A$ and $b_X$ assuming Halofit model with $\sigma_8=0.8$.  ACT fits to $w_{gg}$  and $w_{g\kappa}$ over the 3 points with  $80<l<150$ give $\chi^2=0.64$  and  $\chi^2=0.87$. Planck SGC+NGC fits to $w_{gg}$  and $w_{g\kappa}$ over the 2 points with  $100<l<150$ give $\chi^2=1.47$  and  $\chi^2=1.67$.}
	\label{tab:clgk}
	\begin{tabular}{lcccc} 
Survey     & $b_A$          &  $b_X$         & $b=b_A^2/b_X$  & $\sigma_8'(z=0.15)$  \\
\hline
ACT	       & $1.10\pm0.03$  & $1.10\pm0.15$  & $1.10\pm0.16$  & $0.74\pm0.11$       \\
Planck S+N & $1.10\pm0.05$  & $0.95\pm0.15$  & $1.27\pm0.21$  & $0.64\pm0.10$        \\
Average    & $1.10\pm0.03$  & $1.02\pm0.11$  & $1.16\pm0.13$  & $0.69\pm0.08$        \\
\hline
	\end{tabular}
\end{table}

\begin{table}
	\centering
	\caption{Fits to $17<r<21$ galaxy $w_{gg}$ and $w_{gk}$ for bias $b_A$ and $b_X$ assuming Halofit model with $\sigma_8=0.8$. All fits are made in the range  $20'<\theta<200'$ with $w_{gg}$ giving $\chi^2=1.04$ over 6 points, ACT $w_{gk}$ giving $\chi^2=28.33$ over 12 points and Planck SGC+NGC $w_{gk}$ giving $\chi^2=26.95$ over 12 points.}
	\label{tab:wgk}
	\begin{tabular}{lcccc} 
Survey     & $b_A$          &  $b_X$         & $b=b_A^2/b_X$  & $\sigma_8'(z=0.15)$  \\
\hline
ACT	       & $1.05\pm0.03$  & $1.00\pm0.10$   & $1.10\pm0.12$  & $0.70\pm0.11$       \\
Planck S+N & $1.05\pm0.03$  & $0.83\pm0.08$   & $1.33\pm0.10$  & $0.58\pm0.09$        \\
Average    & $1.05\pm0.02$  & $0.90\pm0.06$   & $1.24\pm0.08$  & $0.63\pm0.07$        \\
\hline
	\end{tabular}
\end{table}
\begin{table}
	\centering
    \caption{Fits to LRG $w_{gg}$ and $C_{g\kappa}^l$ for bias $b_A$ and $b_X$  assuming Halofit model with $\sigma_8=0.8$. Details for $w_{gg}$ are given in Table \ref{tab:wgg_wgk_lrg}. ACT fits to $C_{g\kappa}^l$ over 8 points with  $40<l<260$ give $\chi^2=10.50$. Planck SGC+NGC fits to $C_{g\kappa}^l$ over 4 points with  $100<l<260$ give $\chi^2=6.64$.}
	\label{tab:clgk_lrg}
	\begin{tabular}{lcccc} 
Survey     & $b_A$          &  $b_X$        & $b=b_A^2/b_X$  & $\sigma_8'(z=0.26)$  \\
\hline
ACT	       & $2.18\pm0.03$  & $2.18\pm0.35$ & $1.00\pm0.35$  & $0.67\pm0.35$       \\
Planck S+N & $2.18\pm0.03$  & $1.60\pm0.30$ & $2.97\pm0.30$  & $0.49\pm0.30$        \\
Average    & $2.18\pm0.03$  & $1.89\pm0.23$ & $1.98\pm0.23$  & $0.58\pm0.23$        \\
\hline
	\end{tabular}
\end{table}

\begin{table}
	\centering
    \caption{Fits to LRG $w_{gg}$ and $w_{g\kappa}$ for bias $b_A$ and $b_X$ assuming Halofit model with $\sigma_8=0.8$. The $w_{gg}$ fit over 3 points in the range  $10'<\theta<120'$ gives $\chi^2=3.06$. The ACT $w_{g\kappa}$ fit (with $40<l<3000$ filter) gives $\chi^2=27/52$ over 13 points in the range $10'<\theta<120'$   while the Planck SGC+NGC $w_{g\kappa}$ fit over 12 points in the same range gives $\chi^2=32.39$.}
	\label{tab:wgg_wgk_lrg}
	\begin{tabular}{lcccc} 
Survey     & $b_A$          &  $b_X$         & $b=b_A^2/b_X$  & $\sigma_8'(z=0.26)$  \\
\hline
ACT	       & $2.18\pm0.03$  & $1.9\pm0.17$    & $2.50\pm0.18$  & $0.58\pm0.17$       \\
Planck S+N & $2.18\pm0.03$  & $1.3\pm0.15$    & $3.66\pm0.16$  & $0.40\pm0.15$        \\
Average    & $2.18\pm0.03$  & $1.6\pm0.11$    & $2.97\pm0.12$  & $0.49\pm0.11$        \\
\hline
	\end{tabular}
\end{table}

\subsection{Halofit bias estimates for $0.16<z<0.26$ ATLAS LRGs}
Since the LRG sky density is again quite low, this means that the choice of NSIDE=1024 gives mainly 1's and zeroes in the LRG sky density map with its $\approx4'$ healpix pixel size. Decreasing NSIDE to 512 or 256 gives pixels of 8-16$'$ although fixing the density map issue, compromises the resolution at scales where we have our best data. So we choose to model the $0.16<z<0.36$ LRGs replacing $C_{gg}^l$ by  $w_{gg}$ using the combination of $w_{gg}$ and $C_{g\kappa}^l$ and then $w_{gg}$ and $w_{g\kappa}$ as for the $17<r<21$ galaxies. 

\begin{figure}
	\includegraphics[width=\columnwidth]{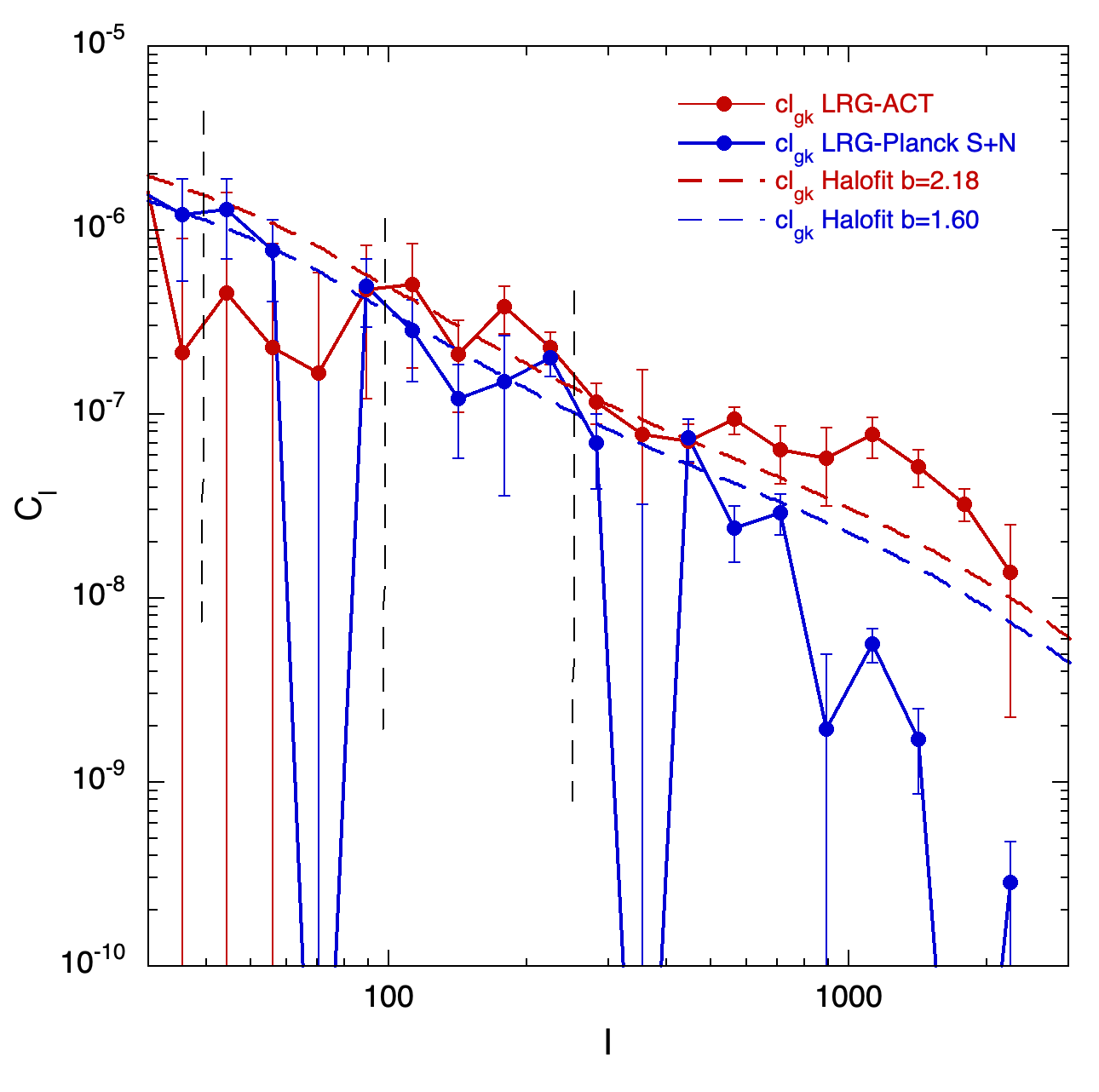}
    \caption{LRG  $C_{g\kappa}^l$ estimates for LRG cross-correlated with ACT and the weighted average of Planck S+N  lensing maps. These are compared to models  based on HALOFIT power spectra from CCL. The vertical dashed lines mark the 2-halo regime fitted for ACT and  include the 8 points in  the $40<l<260$ range. Due to map lensing systematics, the Planck S+N result is only fitted to the 4 points in the range $100<l<260$. }
    \label{fig:wmean-all-halofit}
\end{figure}

\begin{figure}
	\includegraphics[width=\columnwidth]{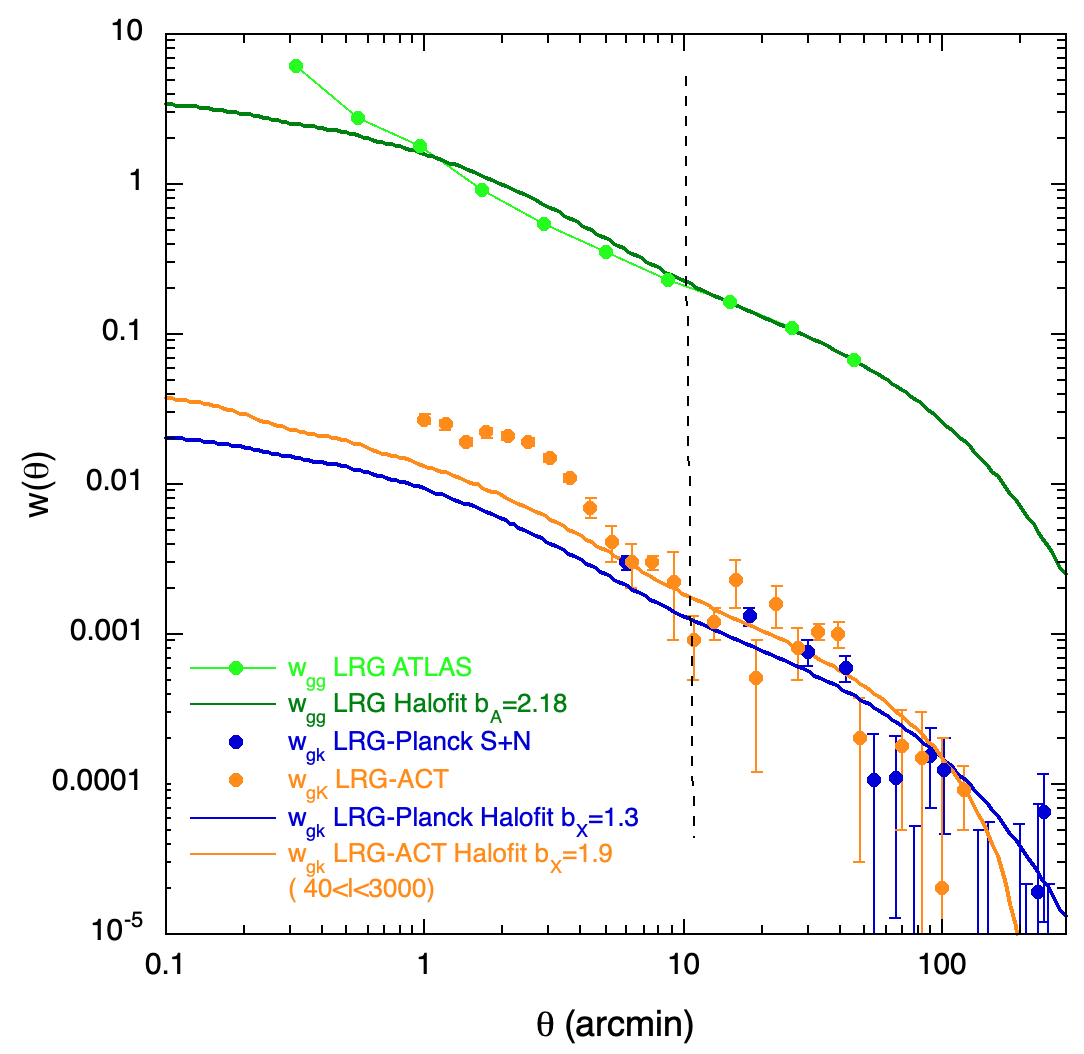}
    \caption{Upper: LRG  $w_{gg}(\theta)$ fitted in range $\theta>10'$ by Halofit model with $b_A=2.18\pm0.03$. Lower: LRG-ACT $w_{g\kappa}(\theta)$ fitted in range $10'<\theta<120'$ by Halofit model ($40<l<3000$) with  $b_X=1.9\pm0.17$; LRG-Planck (SGC+NGC) $w_{g\kappa}(\theta)$ fitted in range $10'<\theta<120'$ by Halofit model with $b_X=1.3\pm0.15$. (See Table \ref{tab:wgg_wgk_lrg}).}
    \label{fig:wggwgk_LRG}
\end{figure}

So $C_{g\kappa}^l$ is shown in Fig. \ref{fig:wmean-all-halofit} and $w_{gg}$ and $w_{g\kappa}$ in Fig. \ref{fig:wggwgk_LRG}. In Fig. \ref{fig:wmean-all-halofit} the results for $C_{g\kappa}^l$ are seen for both ACT and the weighted mean of the Planck SGC and NGC results. Again we prefer not to combine the ACT and Planck results because of the different resolutions of these data. The dashed lines show the range where we will fit our results. We have increased the range to $l=260$ from l=150 in approximation to the average comoving distances between the $17<r<21$ galaxies at $z\approx0.15$ and the LRGs at $z\approx0.26$. At larger scales we maintain the same $l>40$ and $l>100$ limits for ACT and Planck which are set by the scale of systematics in the lensing map data. We fit a Halofit  model assuming $\sigma_8=0.8$ to the 8 ACT points in the range $40<l<260$ and to the 4 Planck points in the range $100<l<260$. In both cases the lower $l$ (large-scale) limit is set by the scales of systematics in the respective lensing maps. However, note that the extra 4 points fitted for ACT at $40<l<100$ have large errors and contribute little to the fit. From Table \ref{tab:clgk_lrg} we see that these fits give $b_X=2.18\pm0.35$ for ACT and $b_X=1.60\pm0.30$ for Planck S+N, the error weighted average of these two giving $b=1.89\pm0.23$. The errors are slightly larger on the ACT result because it only covers the SGC and Planck covers both SGC and NGC ATLAS areas. Also the higher resolution of the ACT data is no help here because we are restricted to the 2-halo regime for the Halofit bias fits. As we shall see below, the LRG $w_{gg}$ fit gives $b_A=2.18\pm0.03$. Note that the errors on the biases ignore covariance between the $C_{gk}^l$ points. This is probably reasonable in this case since $C^l$ modes are uncorrelated in the all-sky limit and the low LRG sky densities also mean that independent Poisson errors dominate.  As summarised in Table \ref{tab:clgk_lrg}, these values for $b_A$ and $b_X$ lead to the final LRG bias estimates of $b=1.0\pm0.35$ with $\sigma_8'(z=0.26)=0.67\pm0.35$ for ACT  and $b=2.97\pm0.30$ with $\sigma_8'(z=0.26)=0.49\pm0.30$ for Planck S+N. Note that here, to estimate $\sigma_8'(z=0.26)$, we are assuming the $\Lambda$CDM growth factor of $D(z=0.26)=0.83$. 

Clearly, there is a significant difference between the $b_X$'s estimated from ACT and Planck S+N and we are unsure of the reason for this. However, we note that here we are using the SGC+NGC  average for $w_{gg}$ and the NGC, only used by Planck, has only a slightly lower $w_{gg}$ than the SGC. Anyway, taking the weighted average of NGC and SGC for the ACT and Planck  $C_{g\kappa}^l$ gives an overall LRG bias estimate of $b=1.98\pm0.23$ with $\sigma_8'(z=0.26)=0.58\pm0.23$. Using the same growth factor implies that $\sigma_8(z=0)=0.70\pm0.28$, statistically consistent with our initial value of $\sigma_8(z=0)=0.8$.

We next compare the above results to those where we replace $C_{g\kappa}^l$  by $w_{g\kappa}$ and combine this LRG angular auto-correlation function with the LRG angular cross-correlation function to estimate the bias $b(z=0.26)$ and $\sigma_8$ at $z=0.26$ and $z=0$. The results for $w_{gg}$ and $w_{g\kappa}$ are shown in Fig. \ref{fig:wggwgk_LRG} and the results of fitting the Halofit model for $b_A$, $b_X$ and the implied LRG bias, $b$, are shown in Table \ref{fig:wggwgk}. We fit at scales $\theta>10'$, represented by the vertical dashed line in Fig. \ref{fig:wggwgk_LRG}, and so again in the linear regime, since this corresponds to $r_{com}>2.3$h$^{-1}$ Mpc at $z=0.26$. We see that $w_{gg}$ is  well fitted by the Halofit model over the 3 points at $\theta>10'$, giving $\chi^2=3.06$ for $b_A=2.18\pm0.03$, although it does less well at smaller scales in the 1-halo regime. The Halofit model also gives a good fit to $w_{g\kappa}$ with bias $b_X=1.9\pm0.17$ for ACT and $b_X=1.3\pm0.15$ for S+N Planck. These values for $b_A$ and $b_X$ give $b=2.5\pm0.18$ for ACT and $b=3.66\pm0.16$ for Planck S+N leading to an overall average of $b=2.97\pm0.12$ for the LRGs, somewhat larger than the $b=1.98\pm0.23$ found using the LRG $C_{g\kappa}^l$. The weighted average between these two estimates of $b(z=0.26)=2.76\pm0.11$ for LRGs gives $\sigma_8'(z=0.26)=0.52\pm0.12$ and assuming the $\Lambda$CDM growth rate gives $\sigma_8'(z=0.0)=0.63\pm0.15$. The $z=0.26$ result for $\sigma_8'$ is compared to the $\Lambda$CDM prediction in Fig. \ref{fig:sigma8_z} and the  results are discussed in Section \ref{sec:discussion}.


\begin{figure}
     \includegraphics[width=\columnwidth]{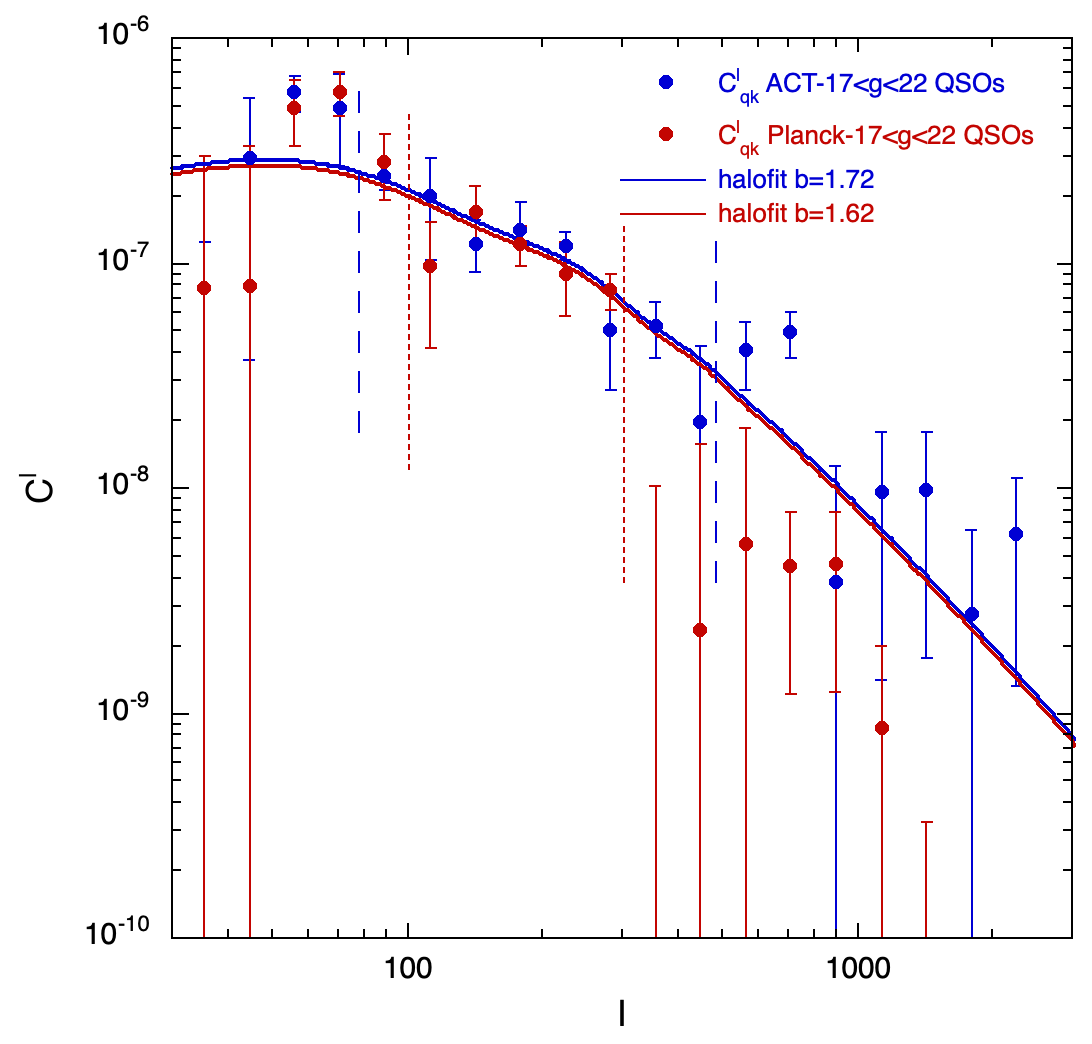}     
    \caption{$C_{qk}^l$  cross-power spectra results for ACT and Planck SGC+NGC lensing maps cross-correlated with  ATLAS $17<g<22$ QSOs. The  models are based on HALOFIT power spectra from CCL, normalised to the best fit values of $b_X$. Blue long- and red short-dashed vertical lines mark the Halofit fitting ranges for $b_X$ for ACT and Planck respectively.  }
    \label{fig:clqk}
\end{figure}

\begin{figure}
    \includegraphics[width=\columnwidth]{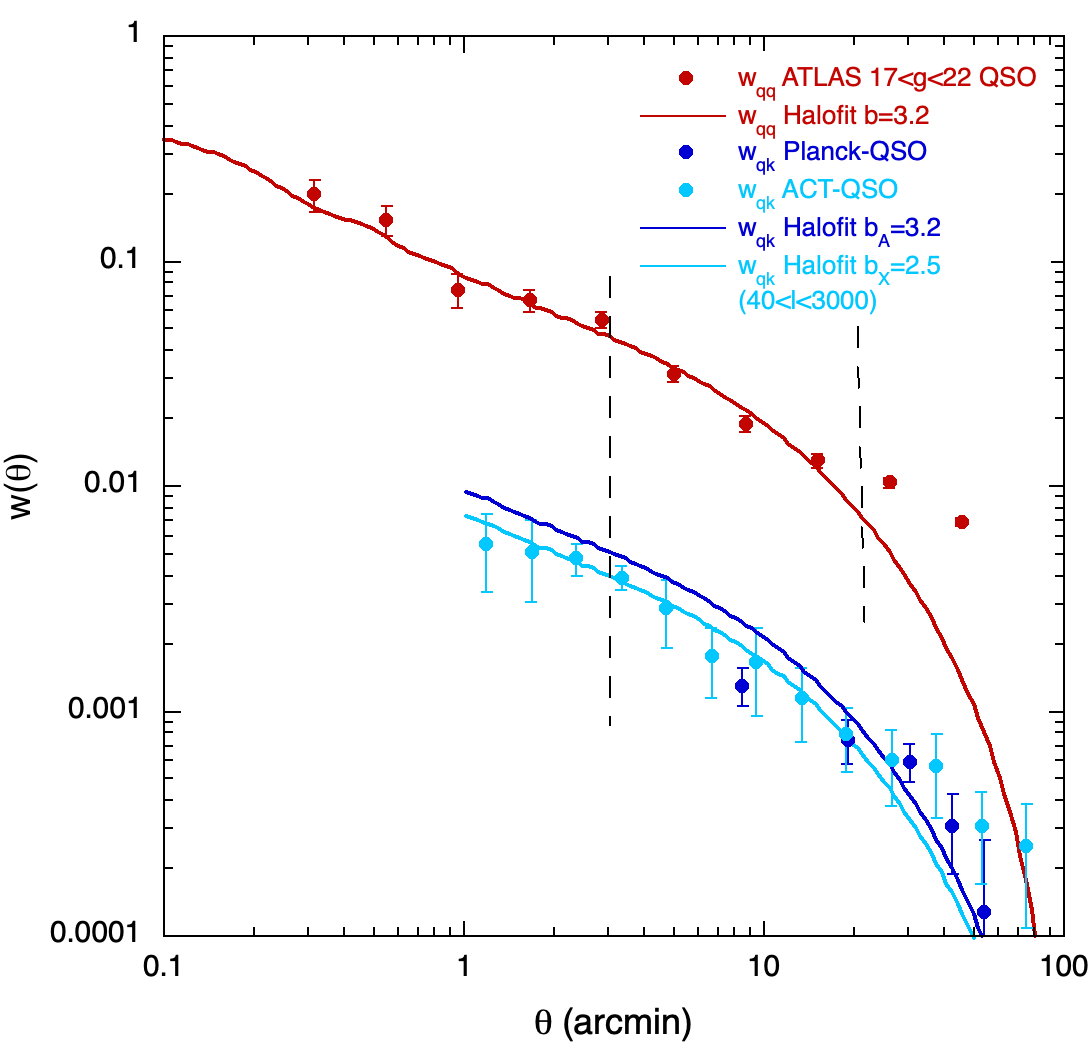}
    \caption{The quasar auto- and cross-correlation functions for $17<g<22$ QSOs, $w_{qq}$ and $w_{qk}$. $w_{qq}$ and the Planck S+N $w_{qk}$ is taken from Eltvedt et al(2024) while the ACT $w_{qk}$ is presented here for the first time. The  models are based on Halofit power spectra from CCL. The ACT best fit model has been pre-filtered by a top-hat filter in the range $40<l<3000$.}
    \label{fig:wqqwqk}
\end{figure}

\begin{figure}
	\includegraphics[width=\columnwidth]{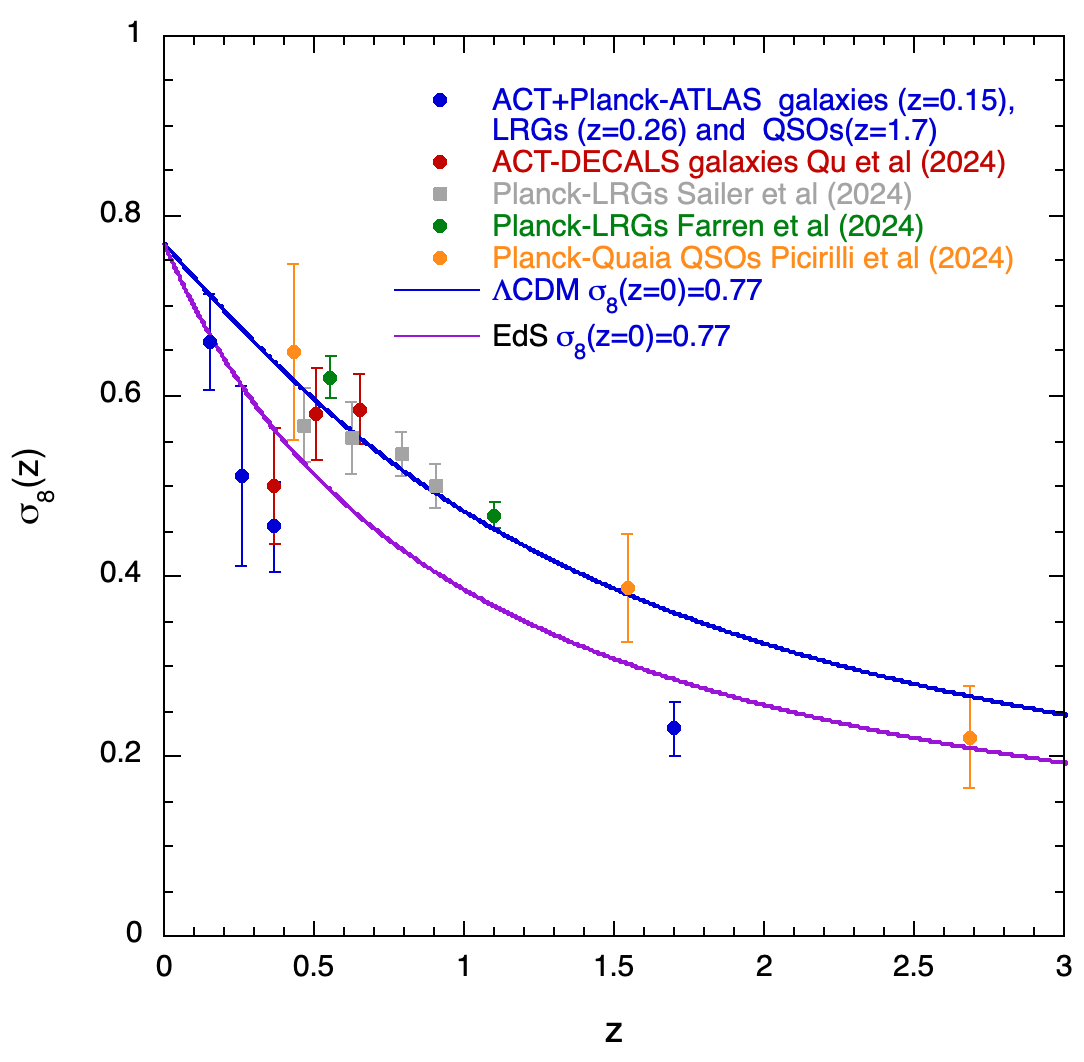}
    \caption{Here we plot $\sigma_8$ (assuming $\Omega_m=0.3$ throughout) versus redshift, z. The ATLAS results are shown as blue points for 17<r<21 galaxies at $z=0.15$, LRGs at z=0.26 and 17<g<22 QSOs at z=1.7 while the blue point at z=0.37 represents our re-analysis of the \protect\cite{Qu_F_2025} result. Also shown are previous results for ACT-DECaLS galaxies (red) in 3 redshift ranges \citep{Qu_F_2025}, Planck-LRGs (grey)  in 4 + 2 redshift ranges out to $z\approx1$ from \protect\cite{sailer2025} (grey) and \protect\cite{Farren2024}(green) and finally QSOs in three redshift ranges from \protect\cite{Pic2024} (orange). Also shown as the blue line is the $\Lambda$CDM growth rate plotted against redshift. Since this growth rate relation appears to decline too slowly with redshift we also show, for comparison, the growth rate - z relation for an Einstein-de Sitter model with $\Omega_m=1$. The faster decline with redshift of this model means it gets closer to fitting the ATLAS $17<g<22$ QSO result at z=1.7.}
    \label{fig:sigma8_z}
\end{figure}


\subsection{Halofit bias estimates for \protect{$z\approx1.7$} ATLAS QSOs.}
\label{sec:halofit-qso}

We first take the same  halofit+bias approach for these $z\approx1.7$ QSOs as we did for the $z\approx0.15$ galaxies. The main difference  is that here, similar to the LRGs, we use $w_{qq}$ rather than $C_{qq}^l$ because the latter statistic has a large Poisson noise contribution to the angular power spectrum at large $l$ due to the low QSO sky density.
Although a Poisson noise term could have been added,  we took the alternative route of analysing $w_{qq}(\theta)$ since the QSO 2-point clustering errors are reasonably Poisson, implying low covariance between the $w_{qq}(\theta)$ points  (see \citealt{BFS1988}).


We therefore proceed to estimate $C_{q\kappa}^l$  for ACT and Planck S+N (see Fig. \ref{fig:clqk}) using the same method and parameters (NSIDE=1024 etc) as for the $17<r<21$ galaxies. In fitting the Halofit model, we take the $80<l<500$ range for ACT and $100<l<300$ for Planck S+N. The ACT upper limit of $l<500$ now corresponds to $r_{com}\approx16$h$^{-1}$Mpc assuming $r_{com}=3300$h$^{-1}$Mpc at $z=1.7$ and so well into the 2-halo regime. The Planck upper limit of $l<300$ for Planck corresponds to  $\approx27'$, just outside the range affected by the Planck beam + $16'$ smoothing. The lower $l$ limits were set as before by the limits imposed by the lensing maps with $l=80$ for ACT corresponding to $\approx100$h$^{-1}$Mpc and $l=100$ for Planck to $\approx80$h$^{-1}$Mpc. Fitting the ACT $C_{q\kappa}^l$ we found  $b_X=1.72\pm0.17$, giving $\chi^2=4.63$ over the 8 points in the range $80<l<500$. Then fitting the Planck S+N $C_{q\kappa}^l$ we found $b_X=1.62\pm0.20$, giving $\chi^2=2.62$ over the 5 points in the range $100<l<300$ (see Table \ref{tab:wqqclqk}).


For our  Halofit fits to $w_{qq}(\theta)$  we used the range $\theta<20'$ corresponding to the range $r_{com}<20$h$^{-1}$ Mpc, (see Fig. \ref{fig:wqqwqk}) which is mostly in the linear regime at this redshift. We cut the range at $\theta<20'$ since the Halofit model is a poor fit to the data at larger scales since the observed $w_{qq}$ continues as a power-law while the model turns over at this point. We found that $b_A=3.2\pm0.04$ ($\chi^2=8.9$, 8 points) was the best fit for $w_{qq}$ in this range.

Our estimates of $w_{q\kappa}$ for the ACT and Planck S+N lensing maps cross-correlated with the $17<g<22$ QSO sample is shown in Fig. \ref{fig:wqqwqk}. Here the ACT results are new whereas the Planck S+N results were taken from \citetalias{Eltvedt_3}. We fitted the Halofit model to the 10 ACT points in the range $3'<\theta<100'$ corresponding to the range $3<r_{com}<20$h$^{-1}$Mpc, which is plausibly  in the linear regime at $z=1.7$ and found a best fit of $b_X=2.5\pm0.20$ with $\chi^2=4.63$. In this ACT case, the Halofit $w_{qk}$ model was pre-filtered in the range $40<l<3000$ before fitting. For  the Planck case, the Halofit model  was fitted to the 7 $w_{q\kappa}$ points in the range $18'<\theta<100'$ with the lower limit taking into account the lower resolution of the Planck data and a best fit of $b_X=3.2\pm0.40$ was found with $\chi^2=2.62$. All the Halofit fitted values are summarised  in Table \ref{tab:wqqwqk} and models assuming these values of $b_A$ and $b_X$ are compared to the observed results in Fig. \ref{fig:wqqwqk}. We note that the ACT and Planck $b_X$ estimates from $w_{q\kappa}$ are both higher than those from $C_{q\kappa}^l$. However, we note that the $b_X$ that show the biggest discrepancies, the Planck S+N, also have the biggest errors in $b_X$ and at least these will be downweighted in obtaining the average bias, b, and $\sigma'_8(z=1.7)$

Here we are assuming $\sigma_8(z=0)=0.8$ and $\Lambda$CDM growth rates in both cases. With these assumptions,  $\sigma_8(z=1.7)=0.8*0.4665=0.373$.  Again $P_{qq}/P_{q\kappa}=b$ implies that $b_Q=b_A^2/b_X$ and so here $b_Q=3.2^2/1.72=5.95$. Similarly, $\sigma_8'(z=1.7)=\sigma_8(z=1.7)*b_X/b_A=0.20\pm0.02$ (see Tables \ref{tab:wqqclqk} and \ref{tab:wqqwqk}). Taking the error-weighted averages, we obtain our final estimates of  $b_Q=4.44\pm0.24$ and $\sigma_8'(z=1.7)=0.23\pm0.03$. We compare this result for $\sigma_8'(z=1.7)$ to the $\Lambda$CDM prediction in Fig. \ref{fig:sigma8_z} and discuss our conclusions in Section \ref{sec:conclusions}.


\begin{table}
	\centering
	\caption{Best Halofit model bias fits to the $17<g<22$ quasar auto-correlation function, $w_{qq}$,  and the  angular cross-power spectra, $C_{qk}^l$, between the  quasars and the  ACT and Planck lensing maps. $w_{qq}$ is fitted for $b_A$ in the range  $\theta<20'$ over  8 points, giving $\chi^2=11.67$ and the  ACT $C_{q\kappa}^l$ for $b_X$ in the range  $80<l<500$ over 8 points for $\chi^2=4.63$. The Planck S+N $C_{qk}^l$ is fitted for $b_X$ in the range  $100<l<300$ over 5 points, giving $\chi^2=2.62$.}
	\begin{tabular}{lcccc} 
Survey      & $b_A$          &  $b_X$           & $b=b_A^2/b_X$  & $\sigma_8'(z=1.7)$ \\  %
\hline
ACT SGC     & $3.20\pm0.04$  & $1.72\pm0.17$    & $5.95\pm0.61$ & $0.20\pm0.02$    \\     
Planck S+N  & $3.20\pm0.04$  & $1.62\pm0.20$    & $6.32\pm0.80$ & $0.19\pm0.02$    \\     
Average     & $3.20\pm0.03$  & $1.68\pm0.13$    & $6.43\pm0.51$ & $0.19\pm0.015$   \\     
\hline
	\end{tabular}
\label{tab:wqqclqk}
\end{table}

\begin{table}
	\centering
	\caption{Best Halofit model bias fits to the $17<g<22$ quasar auto-correlation function, $w_{qq}$, (see Table \ref{tab:wqqclqk} for $b_A$ fit details) and the cross-correlation functions, $w_{qk}$, with the ACT and Planck S+N lensing maps. The ACT  $w_{qk}$ is fitted for $b_X$ by a pre-filtered Halofit model  in the range  $3'<\theta<100'$ over 10 points, giving $\chi^2=8.22$.  The Planck S+N $w_{qk}$ is fitted for $b_X$ in the range $15'<\theta<100'$ over  7 points, giving $\chi^2=5.78$. }
	\label{tab:wqqwqk}
	\begin{tabular}{lcccc} 
Survey      & $b_A$          &  $b_X$           & $b=b_A^2/b_X$  & $\sigma_8'(z=1.7)$ \\
\hline
ACT SGC     & $3.20\pm0.04$  & $2.50\pm0.20$     & $4.10\pm0.34$ & $0.29\pm0.02$      \\       
Planck S+N  & $3.20\pm0.04$  & $3.20\pm0.40$     & $3.20\pm0.41$ & $0.37\pm0.03$      \\       
Average     & $3.20\pm0.03$  & $2.64\pm0.18$     & $3.88\pm0.27$ & $0.31\pm0.02$        \\     
\hline
	\end{tabular}
\end{table}

\section{HODs and non-linear bias for ATLAS galaxies, LRGs and  QSOs.}
\label{sec:hod_nlb_wi98}
Having estimated galaxy, LRG  and QSO bias as a route to the approximate halo masses galaxies and QSOs as well as the cosmological parameter $\sigma_8$, we now turn to fit HOD model parameters directly to these same samples to estimate the galaxy, LRG and QSO halo mass functions in more detail. Since we are now looking to fit non-linear as well as linear scales, we only fit cross-correlations with ACT  and not Planck lensing maps. Since for LRG and QSO auto-correlations we already  only fit angular correlation functions and not angular power spectra, for consistency we shall do similarly for auto- and cross-correlations of all three samples. All   cross-correlation function models will be pre-filtered by the usual $40<l<3000$ top-hat filter.

\subsection{HOD parameter estimates for \protect{$17<r<21$} ATLAS galaxies}

\begin{figure}
    \includegraphics[width=\columnwidth]{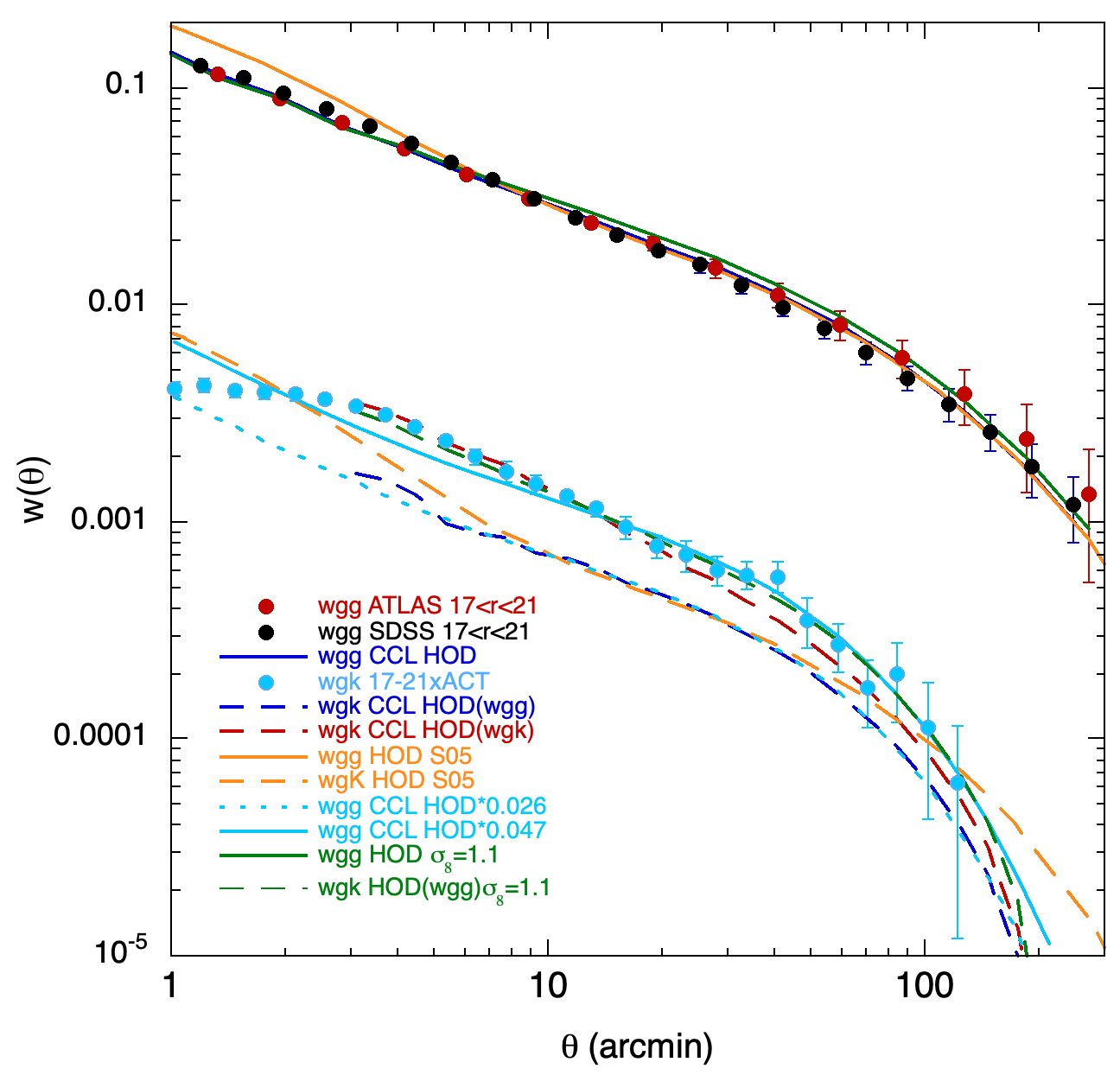}
    \caption{The 17<r<21 galaxy $w_{gg}$ from VST ATLAS (filled red circles) and SDSS (black filled circles) and the 17<r<21 galaxy-ACT $w_{gk}$ results (filled cyan circles) compared to predictions now based on CCL HOD models.  The solid blue line is the HOD fit to $w_{gg}$ with parameters $logM_{min}=13.63$, $logM_0=10.03$, ${logM_1}'=14.12$, $\alpha=1.39$, $\sigma_{lnM}=3.34$ with $\chi^2 = 43.55$ from fitting the 20 points in the range $\theta<300'$.  The dashed blue line is the prediction of this HOD model for $w_{gk}$ - it was not possible to find HOD parameters which simultaneously fitted both $w_{gg}$ and $w_{gk}$. The dashed cyan line is the HOD fit to $w_{g\kappa}$ with parameters $logM_{min}=13.01$, $logM_0=11.09$, ${logM_1}'=14.36$, $\alpha=3.87$, $\sigma_{lnM}=2.53$ with $\chi^2 = 37.09$ from fitting the 23 points in the range $3'<\theta<300'$. Also shown are the models using the  HOD parameters of  \protect\cite{Scranton2005} (orange solid and dashed lines). The HOD models with $\sigma_8(z=0)=1.1$ rather than our fiducial $\sigma_8=0.8$ are shown as the green solid and dashed lines.}
    \label{fig:wgg-wgk-hod}
\end{figure}

\begin{table*}
	\centering
	\caption{Halo Occupation Distribution (HOD) parameters estimated from fitting auto- and cross-correlation functions for galaxies, LRGs and QSOs. The HOD model is from \protect\cite{Z2007}.  In the first column, `predicted' means the model parameters from fitting the auto-correlation function have been assumed in obtaining the $\chi^2$ goodness-of-fit for the cross-correlation function. In the final column, $N$ represents the number of points fitted for the $\chi^2$ values shown. For the QSO HOD fits we have assumed that $\log(M_{min})$ = $\log(M_0)$ and so we only fit 4 parameters in these cases.}
	\label{tab:hodfits}
	\begin{tabular}{lcccccc} 
Survey                                               & $\log(M_{min})$ & $\log(M_0)$       &  ${logM_1}'$ & $\alpha$       & $\sigma_{lnM}$  & $\chi^2$(N)\\
\hline
ATLAS $17<r<21$ galaxies $w_{gg}$ fit                & $13.63\pm0.04$  & $10.03\pm0.20$    & $14.12\pm0.34$ & $1.39\pm0.02$ & $3.34\pm0.15$  & 43.55 (19)\\ 
ATLAS $17<r<21$ galaxies $w_{gk}$ fit                & $13.01\pm0.04$  & $11.09\pm0.20$    & $14.36\pm0.34$ & $3.87\pm0.02$ & $2.53\pm0.15$  & 37.09 (24)\\ 
ATLAS $17<r<21$ galaxies $w_{gg}$ + $w_{gk}$ $>3'$   & $14.77\pm0.04$  & $10.89\pm0.20$    & $15.00\pm0.34$ & $1.47\pm0.02$ & $3.64\pm0.15$  & 94.5 (19) 237.5 (24)\\
\hline
ATLAS $17<r<21$ galaxies $w_{gg}$ $\sigma_8=1.1$ fit & $13.69\pm0.04$  & $11.24\pm0.20$    & $13.71\pm0.34$ & $0.94\pm0.02$ & $3.98\pm0.15$  & 60.58 (19)\\ 
ATLAS $17<r<21$ galaxies $w_{gk}$ $\sigma_8=1.1$ (predicted)& $13.69\pm0.04$  & $11.24\pm0.20$    & $13.71\pm0.34$ & $0.94\pm0.02$ & $3.98\pm0.15$  & 39.86(24)\\ 
\hline
ATLAS $0.16<z<0.36$ LRGs $w_{gg}$ fit               & $14.48\pm0.04$  & $11.84\pm0.40$    & $16.60\pm0.41$ & $0.71\pm0.03$ & $1.91\pm 0.01$ & 1567.1 (10)\\        
ATLAS $0.16<z<0.36$ LRGs $w_{gk}$ fit               & $15.85\pm0.04$  & $11.94\pm0.40$    & $15.94\pm0.41$ & $3.34\pm0.03$ & $2.74\pm0.01$ & 39.18 (21) \\
ATLAS $0.16<z<0.36$ LRGs $w_{gk}$ (predicted)        & $14.48\pm0.04$  & $11.84\pm0.40$    & $16.60\pm0.41$ & $0.71\pm0.03$ & $1.91\pm 0.01$ & 71.2 (21)\\ 
ATLAS $0.16<z<0.36$ LRGs $w_{gg}$+$w_{gk}$ fit      & $14.33\pm0.04$  & $11.60\pm0.40$    & $16.99\pm0.41$ & $0.58\pm0.03$ & $1.78\pm0.01$  & 993.6(10) 79.4(20) \\ 
ATLAS $0.16<z<0.36$ LRGs $w_{gg}$+$w_{gk}$ (predicted) & $14.48\pm0.04$  & $11.84\pm0.40$    & $16.60\pm0.41$ & $0.71\pm0.03$ & $1.91\pm 0.01$ & 1567.1(10) 71.2 (21)\\ 
\hline
ATLAS $17<g<22$ QSOs $w_{qq}$ fit $\sigma_8=0.8$ $<20'$ & $13.13\pm0.04$  & $13.13\pm0.40$    & $13.95\pm0.41$ & $2.12\pm0.03$ & $1.26\pm 0.01$ & 11.6(8) \\ 
ATLAS $17<g<22$ QSOs $w_{qk}$ fit $\sigma_8=0.8$ $>3'$ & $14.00\pm0.03$  & $14.00\pm0.18$    & $12.77\pm0.27$ & $2.03\pm0.02$ & $3.59\pm0.02$  & 44.0 (17) \\ 
ATLAS $17<g<22$ QSOs $w_{qq}$+$w_{qk}$ fit $\sigma_8=0.8$& $12.69\pm0.03$& $12.69\pm0.18$ & $13.80\pm0.27$ & $2.22\pm0.02$ & $0.49\pm0.02$  & 12.9 (8) 53.6 (17) \\   
\hline
	\end{tabular}
\end{table*}

\begin{figure}
     \includegraphics[width=\columnwidth]{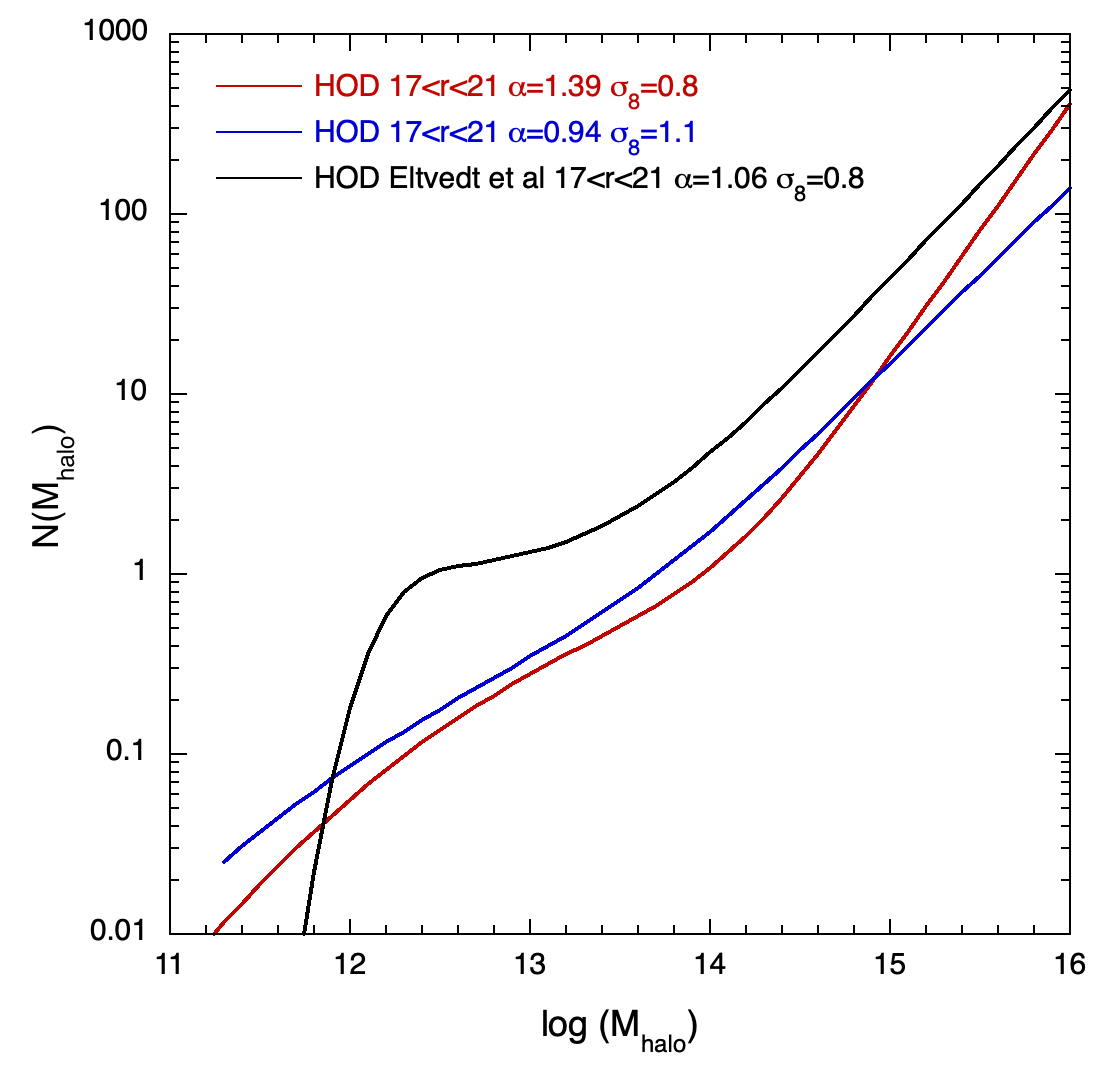}
     \includegraphics[width=\columnwidth]{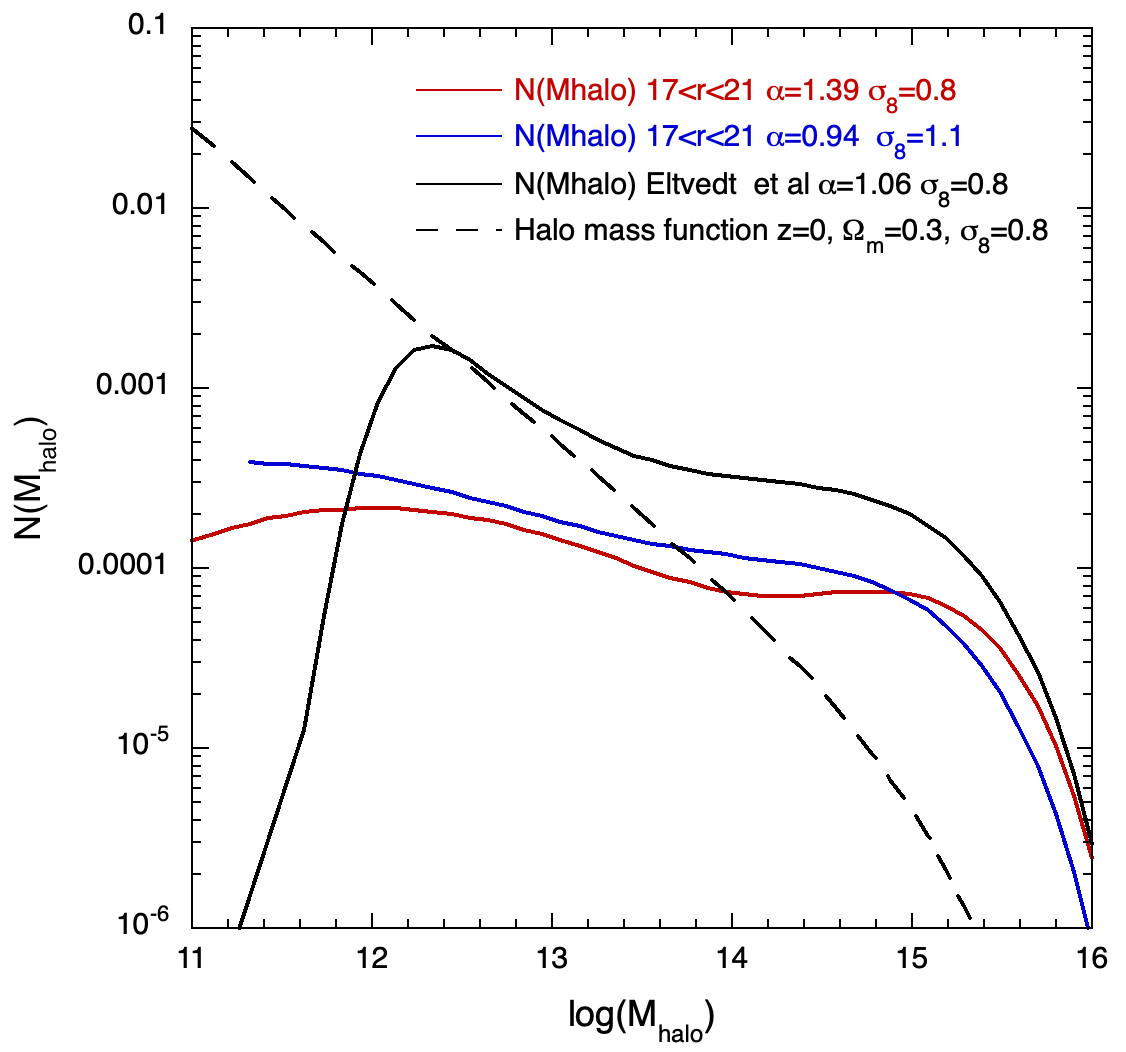}
     \caption{(a-top) HOD models are now shown for the 17<r<21 galaxies based on the fitted parameters in Table \ref{tab:hodfits}. These are  $logM_{min}=13.63$, $logM_0=10.03$, ${logM_1}'=14.12$, $\alpha=1.39$, $\sigma_{lnM}=3.34$ and assuming $\sigma_8=0.8$ (blue line). The  red line is the HOD fit to $w_{gg}$ with parameters $logM_{min}=13.69$, $logM_0=11.24$, ${logM_1}'=13.71$, $\alpha=0.94$, $\sigma_{lnM}=3.98$ with $\sigma_8=1.1$ (see Table \ref{tab:hodfits}). The black line is the HOD model prediction from \protect\citetalias{Eltvedt_2} with $logM_{min}=12.17$, $logM_0=11.53$, ${logM_1}'=13.46$, $\alpha=1.06$, $\sigma_{lnM}=0.60$ and assuming $\sigma_8=0.8$. (b-bottom) The 17<r<21 galaxy N - halo mass distribution, based on the above HOD represented by the red line.}
\label{fig:HOD_Nm_gal}
\end{figure}

Fig. \ref{fig:wgg-wgk-hod}  shows the ACT cross-correlation function for $17<r<21$ galaxies 
and also the auto-correlation function estimated in \citetalias{Eltvedt_2} for this sample, as previously shown in Fig. \ref{fig:wggwgk}. 
These correlation functions are now compared to HOD models, independently MCMC fitted assuming the  5 parameter HOD model of \cite{Z2007}. The best fit parameters to the auto- and cross-correlation functions are given in Table \ref{tab:hodfits} and the caption to Fig. \ref{fig:wgg-wgk-hod}. However, it was not possible to find HOD parameters that simultaneously fitted both auto- and cross-correlation functions. For example, if we assume the best fit auto-correlation parameters, these give the blue dashed line which significantly underestimates the cross-correlation function at $\theta<50'$, although fitting better $w_{g\kappa}$ in the linear regime at larger scales. The same conclusion applies if the cross-correlation fit is assumed and indeed when the MCMC procedure is applied jointly to both results (see Table \ref{tab:hodfits}).  

Then we considered if, similar to the Halofit case, whether self-consistent HOD fits to $w_{gg}$ and $w_{g\kappa}$ could be obtained  by allowing $\sigma_8$ to vary when fitting the HOD parameters. We found that if we assume $\sigma_8(z=0)=1.1$ and then refitted the HOD parameters then  the $w_{g\kappa}$ fit improved significantly (see Fig. \ref{fig:wgg-wgk-hod} and Table \ref{tab:hodfits}), now giving  $\chi^2=39.86$ when applied to $w_{g\kappa}$ over 19 points, an excellent fit compared to  the fit with $\sigma_8(z=0)=0.8$. Taking $b=0.55$ from the WI98 method and assuming $\sigma_8=0.8$ if $b=1$, also results in an even higher value of $\sigma_8=1.45$, with these results overlapping the HOD predicted $w_{gk}$ result with $\sigma_8=1.1$. However, these higher than expected  values for $\sigma_8$ from the non-linear regime are in considerable disagreement with the lower than expected  Halofit values from the linear regime. Clearly $\sigma_8$ could be regarded as another parameter alongside the bias/HOD and $\Omega_m$ that can affect CMB lensing fits in this non-linear regime but the Halofit estimates of $\sigma_8$ from the linear regime are preferred since they are independent of uncertainties in the modelling of the 1-halo term. We note that we could lower s8 to 0.74 and this would increase the antibias but conservatively we maintain the fiducial value of $\sigma_8(z=0)=0.8$ for this non-linear analysis. 



Also shown are the HOD model predictions based on the parameters of \cite{Scranton2005} (S05) that were compared  to the angular galaxy correlation function and the galaxy-Planck cross-correlation function in \citetalias{Eltvedt_2} for this galaxy sample. We see that the same conclusion applies here with the S05 HOD giving a good fit to the auto-correlation function at least in the range $\theta>5'$ but underestimating the cross-correlation function at $3'<\theta<15'$, more than in the other case.

We compare the new HOD fit for the $17<r<21$ galaxies with that of \citetalias{Eltvedt_2} in Fig. \ref{fig:HOD_Nm_gal}(a). We see that the new HOD with $\sigma_8=0.8$ (red line) lies below the previous HOD (black line) and has a much flatter slope up to $log(M_{halo})\approx13$ due to a bigger contribution from $\sigma_{lnM}=3.34$. A similar conclusion broadly applies to the HOD that was fitted assuming $\sigma_8=1.1$. When the halo mass function that applies in both these cases (only the $\sigma_8=0.8$ version is shown as the black dashed line in Fig. \ref{fig:HOD_Nm_gal}(b) is multiplied by the HODs in (a), we see that both the galaxy host halo mass functions are relatively flat between $11<log(M)<15.5$ whereas the previous fit of \citetalias{Eltvedt_2} (blue line) cut off more quickly at $log(M)\approx12.3$. Clearly the new HOD and galaxy mass function is preferred due to the higher resolution of the ACT lensing map producing a better defined $w_{g\kappa}$ at small scales and hence more accurately fitted HOD parameters.

\subsection{Non-linear bias for \protect{$17<r<21$} ATLAS galaxies}
The S05 HOD is interesting because \cite{MS2007} suggested that it fitted the same $17<r<21$ galaxy galaxy-QSO lensing data better than a model where galaxies trace the mass. This  \cite{WI98} model predicts that $w_{g\kappa}=\kappa/b\times w_{gg}$ and several authors had suggested that this model under-predicted the $w_{gq}$ data with standard $\Lambda$CDM $\Omega_m=0.3$ and bias, $b=1$ parameters. \cite{MS2007} concluded that since the observed $w_{gq}$ between S05 and \cite{Myers2005} were almost identical, the S05 standard model fit  must be due to their inclusion of a HOD.  However, in our more recent analysis of $w_{gg}$ and $w_{gq}$ in \citetalias{Eltvedt_2} we find that the S05 model underpredicts the amount of galaxy-QSO lensing in the observed $0.2<\theta<10'$ range and that the $w_{gq}$ data still prefers a WI98 model with anti-bias $b\approx0.5$ (see their Fig. 6 and Fig. 7b,c). And here Fig. \ref{fig:wgg-wgk-hod} shows that the S05 HOD seems to significantly  underpredict the ACT $w_{gk}$ data in this range where there are far smaller errors for the ACT CMB lensing analysis than for the QSO lensing analysis of \citetalias{Eltvedt_2}.

So in detail, Fig. \ref{fig:wgg-wgk-hod} therefore compares  the WI98 and S05 model predictions in this CMB lensing case. The dashed cyan line in Fig. \ref{fig:wgg-wgk-hod} represents the WI98 $w_{gk}$ model obtained by reducing the $w_{gg}$  fit (red line) by a factor of 20 with only the $40<l<3000$ filter added. Since for this sample $\kappa=0.026$ for $0.1<z<0.36$ but, fitting in the range $3'<\theta<140'$, the best fit is $\kappa/b=0.047\pm0.001$ implying $b=0.55\pm0.01$. We note that the cyan line with $\kappa=0.026$ is close to the S05 HOD model in the $0.'5<\theta<10'$ range covered by $w_{gq}$ (see Figs. 6, 7b,c of \citetalias{Eltvedt_2}). Indeed,  anti-bias may be implied out to $\theta\approx 40'$ ($\approx5$h$^{-1}$ Mpc) with $b_g=0.55$, although at larger scales  $b\approx1$. We also conclude that $w_{gq}$ is not well fitted in the $\theta<40'$ range by the S05 HOD. We note the high quality of fit to $w_{g\kappa}$ of the WI98 model with $\kappa=0.046$ which gives $\chi^2=22.51$ over 15 points in the range $12<\theta<200$ arcmin covering both 1- and 2-halo terms. This compares to  the Halofit fit to the ACT $w_{gk}$ which gave $\chi^2=28.33$ over 12 points in the range $20<\theta<200$ arcmin. Thus the WI98 model actually has a better $w_{g\kappa}$ goodness-of-fit with a reduced $\chi^2=1.5$ than Halofit (reduced $\chi^2=2.36$) despite its shorter fitting range. Indeed, the WI98 model reduced $\chi^2=2.27$ when the range fitted is increased to the $6<\theta<200$ arcmin (19 points), still less than Halofit in its $20<\theta<200$ arcmin range.

We also note that Fig. \ref{fig:wggwgk} shows that the Halofit model also seems to give a more self-consistent fit between  $w_{gg}$ and $w_{gk}$ at least in the 2-halo regime than the HOD model. Indeed, we generally conclude that all models are self-consistent in the 2-halo regime but the model where galaxies trace the mass perform better than the other two halo based models in the 1-halo regime. We shall return to discuss these results further in Section \ref{sec:discussion}.

Given the excellent fit of the WI98 model with $b=0.55\pm0.01$, it may be now worth briefly considering whether the degeneracy between the bias $b$ and $\Omega_m$ in these models might imply a higher $\Omega_m$ rather than a low bias. So measuring $b=0.55\pm0.01$ for $\Omega_m=0.3$, implies $\Omega_m=0.3/0.55=0.54\pm0.01$ if we now assume b=1. Indeed, if we take the Halofit estimate of $b=1.20\pm0.07$ rather than b=1 then $\Omega_m=0.3/0.55\times1.20=0.66\pm0.04$. Although this estimate for $\Omega_m$ is in tension with the estimates from SNIa \citep{Riess1998} and BAO \citep{Eisenstein2005} Hubble diagrams, it is at least self-consistent with the high gravitational growth rate suggested  in Fig. \ref{fig:sigma8_z} above. 


Finally, recall that the Halofit results assumed $\Omega_m=0.3$ and gave $b=1.2$, with $\sigma_8\approx0.8$ so here $\Omega_m/b=0.3/1.2=0.25\pm0.02$ rather than $\Omega_m/b=0.66\pm0.04$ which is a highly significant ($\approx9\sigma$) difference. Now if this difference was less significant, an argument might be made that given the better quality of fit  of the WI98 `galaxy tracing mass' model over both non-linear and linear scales than the Halofit model, the WI98 estimate of $\Omega_m/b$ might be preferred to the Halofit model even in the linear regime leading to the higher estimate of $\Omega_m=0.5-0.7$ that this would imply with no puzzle remaining over the anti-bias required if $\Omega_m=0.3$. But given the above tensions implied with the SNIa and BAO Hubble diagrams, more data will be needed before these two indications of a higher $\Omega_m$ from CMB lensing can be taken seriously.

\subsection{HOD parameter estimates for \protect{$z\approx0.26$} ATLAS LRGs}

\begin{figure}
     \includegraphics[width=\columnwidth]{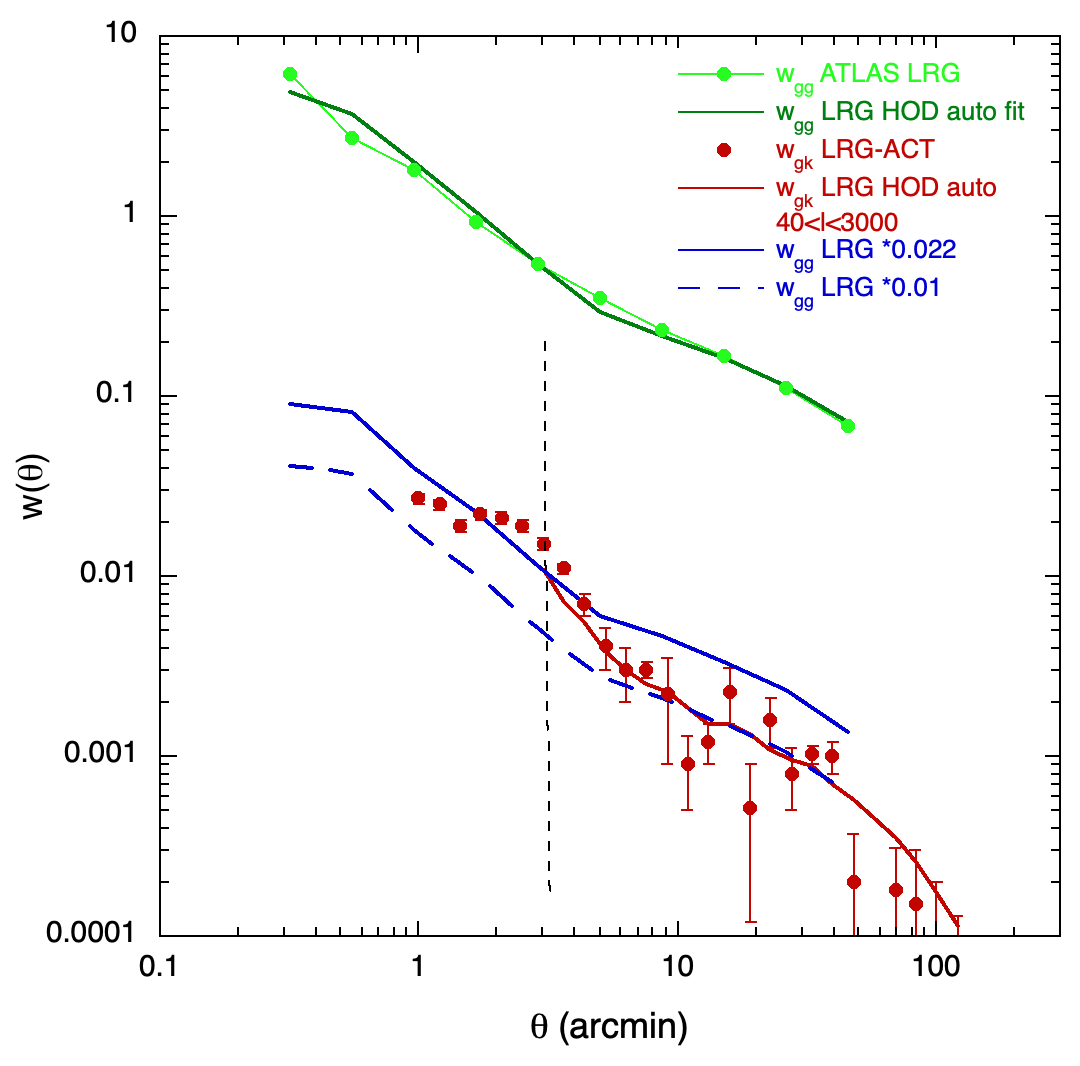}
     \caption{HOD models are now compared to the $w_{gg}$ (green filled circles) and $w_{g\kappa}$ (red filled circles) results for LRGs. The solid green line is the HOD fit to $w_{gg}$ with parameters $logM_{min}=14.48$, $logM_0=11.84$, ${logM_1}'=16.60$, $\alpha=0.71$, $\sigma_{lnM}=1.91$ with $\chi^2 = 1567.1$ from fitting the 10 points in the range $0.'3<\theta<50'$ (see Table \ref{tab:hodfits}). The red line is this same HOD model prediction for $w_{gk}$ with the $40<l<3000$ filter now also  applied. We see that the HOD fit to the LRG $w_{gg}$ when translated to the LRG $w_{g\kappa}$ provides a  self-consistent fit with $\chi^2 = 71.2$ when fitting over 20 points in the range $\theta>3'$. Finally, the dashed and solid blue lines represent the LRG $w_{gg}$ results (here the HOD fit) scaled by a factor of $\kappa/b=0.022$ (where $\kappa=0.022$ and b=1) and 0.01 (fitted for $5'<\theta<45'$) for comparison with $w_{g\kappa}$ in the context of the WI98 model where galaxies are assumed to trace the mass. Both models are now  pre-filtered in the range $40<l<3000$. The $\kappa/b=0.01$ model matches the LRG $w_{g\kappa}$  result at large scales while the other $b=1$ model fits better at smaller scales,  although here the HOD model fit is much more consistent between $w_{gg}$ and $w_{g\kappa}$ than it was for the $17<r<21$ galaxies.}
\label{fig:wggwgkHOD_LRG}
\end{figure}

The LRG $w_{gg}$ (green points) and $w_{g\kappa}$ (red points) estimates are shown in Fig. \ref{fig:wggwgkHOD_LRG}, as previously presented in Fig. \ref{fig:wggwgk_LRG}. We first fit  $w_{gg}$ independently, and $w_{gg}$ + $w_{g\kappa}$  jointly, finding the HOD parameters given in Table \ref{tab:hodfits} and in the caption of Fig. \ref{fig:wggwgkHOD_LRG}. We note that although the $\chi^2$ for $w_{gg}$ is extremely poor, this can be viewed as the small size of the error bars in the LRG $w_{gg}$ coming into conflict with the detailed form of the HOD model but since we are only looking for a HOD within the limits of the \cite{Z2007} parameterisation that forms a broad representation of the observed $w_{gg}$, we consider this fit adequate. The best joint fit to $w_{gg}$ and $w_{g\kappa}$ is the combination of the best fits to these two functions separately, suggesting they are consistent in terms of the \cite{Z2007} HOD model parameterisation. To test this consistency further, we then predict $w_{g\kappa}$, assuming the $w_{gg}$ fit parameters and from Fig. \ref{fig:wggwgkHOD_LRG} and Table \ref{tab:hodfits}. We see that the $w_{gg}$ model gives a reasonably self-consistent fit to the observed $w_{g\kappa}$ in the range $\theta>3'$ i.e. rightwards from the vertical dashed line, with $\Delta\chi^2=8.2$ relative to the best fit. At smaller scales the effect of the ACT $1'-3'$ resolution will need to be accounted for before the quality of fit to the 1-halo term can be judged. However, it does seem like the \cite{Z2007} HOD model may give a more self-consistent  fit to the LRGs than the equivalent HOD model did for the $17<r<21$ galaxies.

\begin{figure}
     \includegraphics[width=\columnwidth]{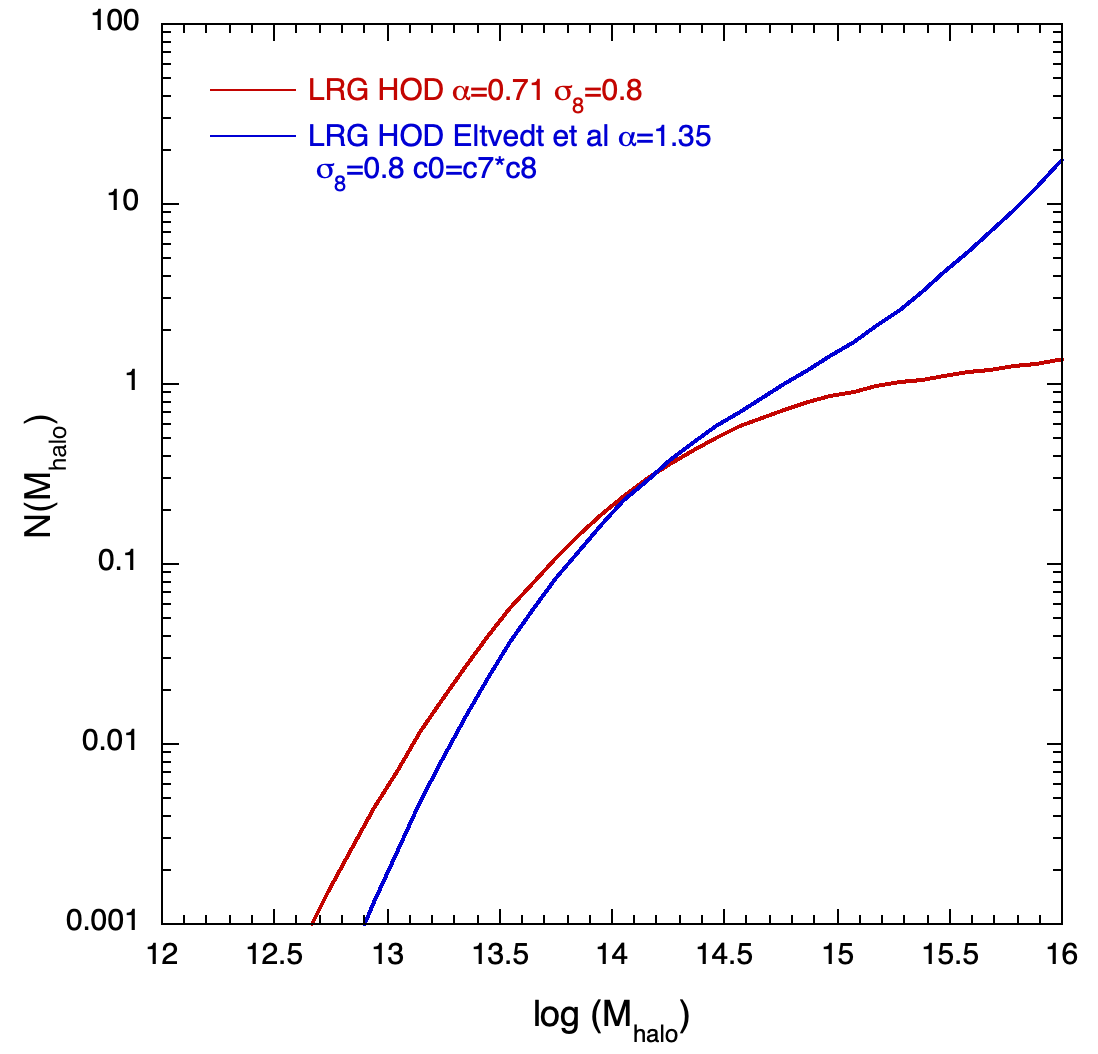}
     \includegraphics[width=\columnwidth]{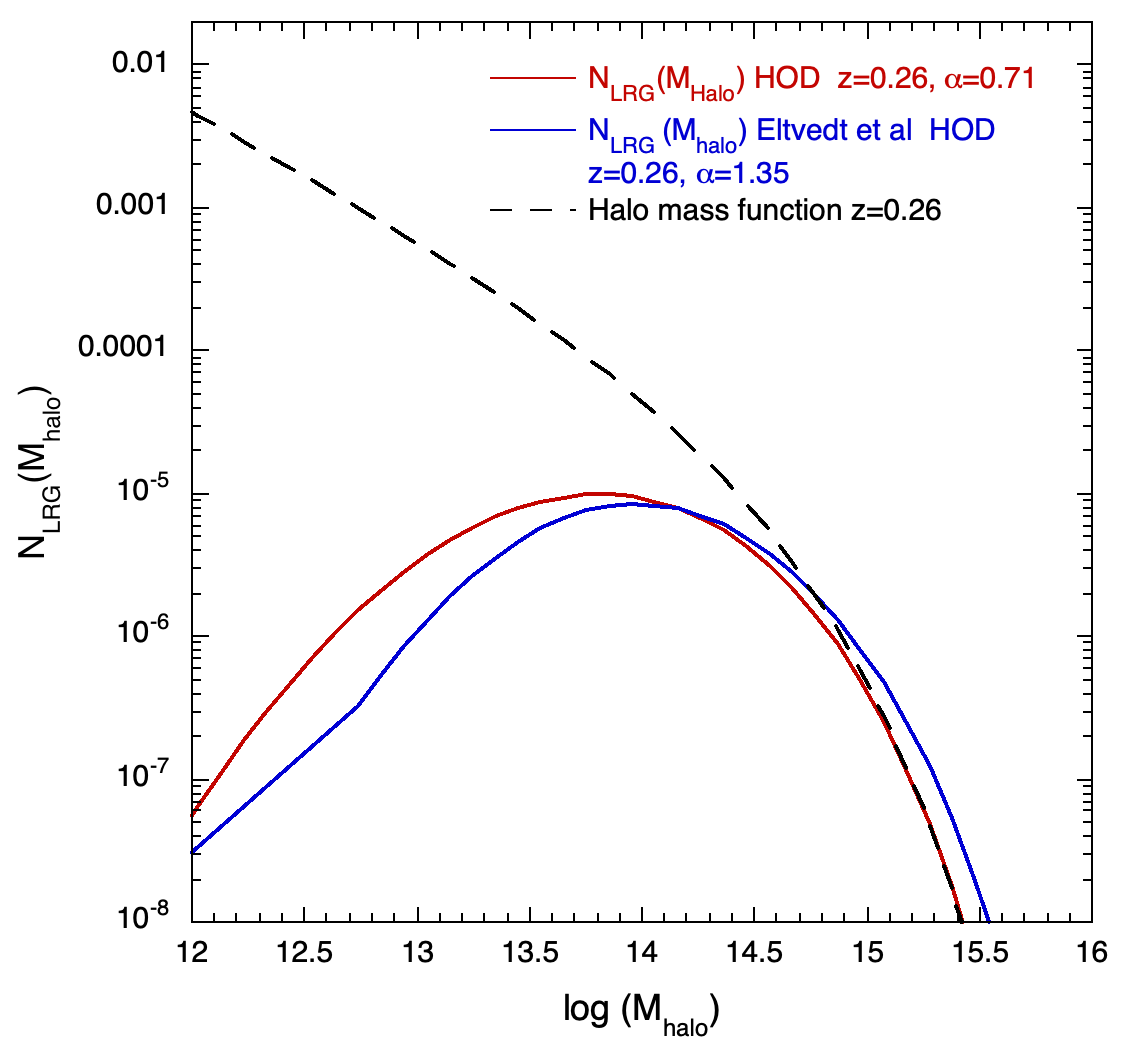}
     \caption{(a-top) HOD models are now shown for the 0.16<z<0.36 LRGs based on the fitted parameters in Table \ref{tab:hodfits}, here always assuming $\sigma_8=0.8$. These are  $logM_{min}=14.48$, $logM_0=11.84$, ${logM_1}'=16.60$, $\alpha=0.71$, $\sigma_{lnM}=1.91$ and assuming $\sigma_8=0.8$ (red line). The blue line is the HOD model prediction from \protect\citetalias{Eltvedt_2} with $logM_{min}=14.45$, $logM_0=12.64$, ${logM_1}'=15.10$, $\alpha=1.35$, $\sigma_{lnM}=1.63$ and assuming $\sigma_8=0.8$. (b-bottom) The $0.16<z<0.36$ galaxy $N_{LRG}$ - halo mass distributions, based on the above HODs represented in  (a-top) and (b-bottom)  by the red and blue lines, for comparison purposes.  }
\label{fig:HOD_Nm_LRG}
\end{figure}

Fig. \ref{fig:HOD_Nm_LRG}(a-top) compares the new LRG  HOD to that found in \citetalias{Eltvedt_2} and Fig. \ref{fig:HOD_Nm_LRG}(b-bottom) compares the LRG halo mass functions found by multiplying the halo mass function by the HODs in (a-top). We first see that the two mass functions in (b-bottom) are similar, peaking at around $log(M)\approx14$ with the new mass function extending to slightly lower masses than the HOD of \citetalias{Eltvedt_2}. As expected, we see that both LRG halo mass functions comprise the highest masses of the halo mass distribution out to log(M)=16.0. They therefore also comprise the highest mass end of the $17<r<21$ galaxy halo mass distribution in Fig. \ref{fig:HOD_Nm_gal}(b-bottom). Again the higher resolution of ACT compared to Planck means that the HOD fits to $w_{g\kappa}$ in this paper supersede those of \citetalias{Eltvedt_2}.

\subsection{Non-linear bias estimates for ATLAS LRGs}

We first note that a model where LRGs trace the mass obtained by scaling $w_{gg}$ downwards to fit $w_{g\kappa}$ gives a  less compelling  fit here than it did in the case of the $17<r<21$ galaxies. The solid blue line in Fig. \ref{fig:wggwgkHOD_LRG} is the LRG $w_{gg}$ HOD fit shifted downwards by $\kappa/b$ with $\kappa=0.022$ calculated for $0.16<z<0.36$ and $b=1$ while the dashed blue line is  $w_{g\kappa}$ with $\kappa/b=0.01\pm0.0002$. With the calculated $\kappa=0.022$, this latter model  implies $b=2.2\pm0.04$ ($\chi^2=60.2$ over 12 points in the range $5`<\theta<45'$).  Clearly, the $b=2.2$ ($\kappa/b=0.01$) model (dashed) fits better at large scales while the $b=1$ ($\kappa/b=0.01$) model(solid) fits better at small scales. These biases are both higher than their equivalents for the $17<g<21$ galaxies where $b=1.2$ was found at large scales and $b\approx0.5$ at small scales. Moreover, because the shape of $w_{gg}$ for the LRGs is a less good match to the LRG $w_{g\kappa}$ than it was for $17<r<21$ galaxies suggests that the LRGs do not trace the mass over the full  range of scales probed here and certainly much less well than the $17<r<21$ galaxies. The same conclusion is reached from the goodness-of-fit of the  HOD model to the LRG $w_{g\kappa}$, consistent with  there  being more mass in the 1-halo regime relative to the 2-halo regime than implied by the LRG $w_{gg}$ again implying  LRG clustering does not trace mass.



\begin{figure}
     \includegraphics[width=\columnwidth]{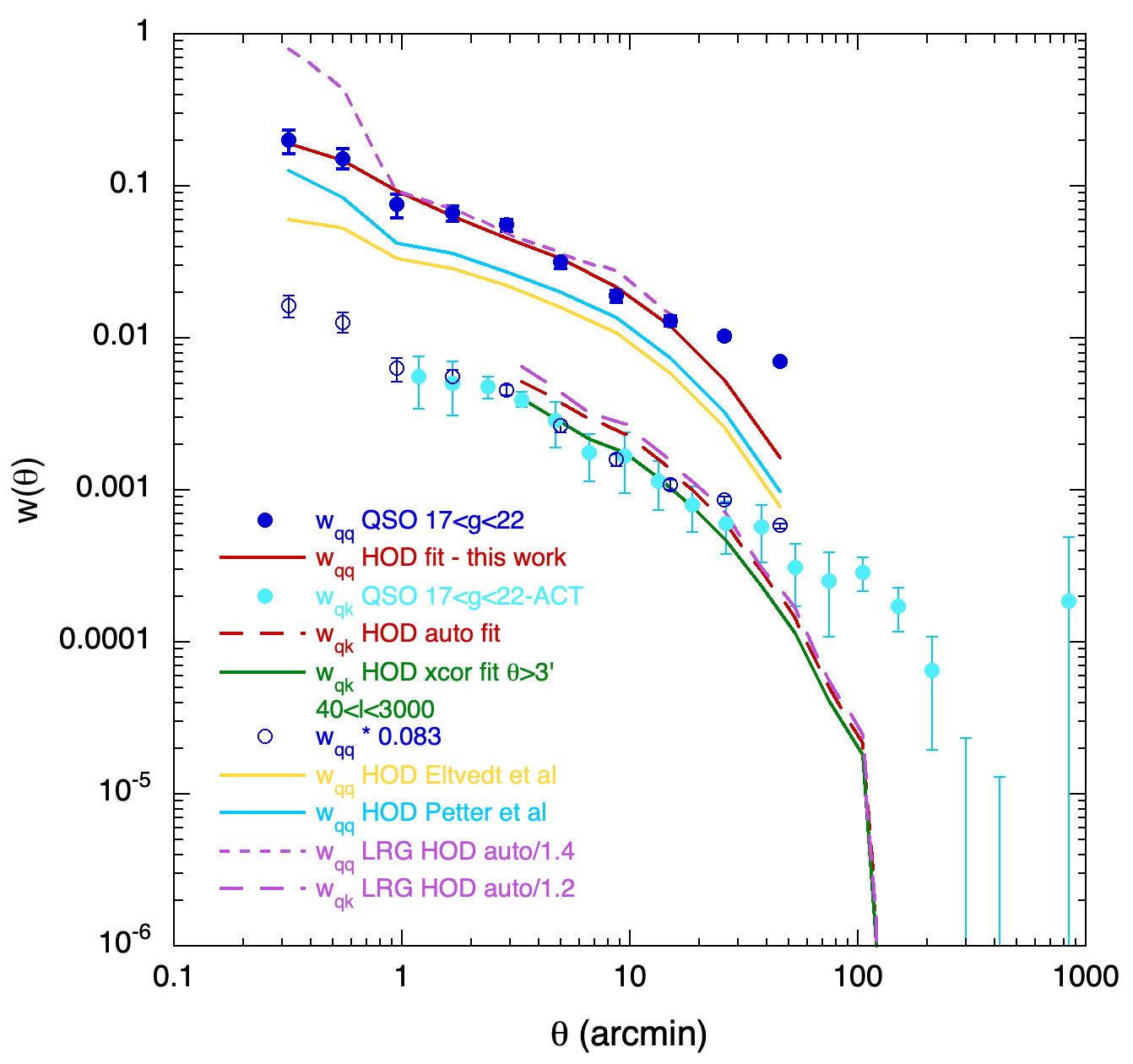}
     \caption{$w_{qq}$ and $w_{q\kappa}$ HOD models are now compared to the $w_{qq}$ and $w_{q\kappa}$ data. The solid red line is the HOD fit to $w_{qq}$ with parameters $logM_{min}=logM_0=13.13$, ${logM_1}'=13.95$, $\alpha= 2.12$, $\sigma_{lnM}=1.26$ with $\chi^2 = 11.6$ fitting the 8 points in the range $0.'3<\theta<20'$. The red dashed line is this HOD model prediction for $w_{q\kappa}$ which gives $\chi^2= 56.7$. The green solid line is the HOD model fit to $w_{q\kappa}$ with parameters $logM_{min}=logM_0=14.00$, ${logM_1}'=12.77$, $\alpha= 2.03$, $\sigma_{lnM}=3.59$ with $\chi^2= 44.0$ fitting the 17 points in the range $\theta>3'$. Note that the $40<l<3000$ filter has been applied to both these $w_{q\kappa}$ models. Moving from $\sigma_8=0.8$ to $\sigma_8=0.6$ gives consistent fits with parameters $logM_{min}=logM_0=13.96$, 
     ${logM_1}'=12.81$, $\alpha=3.89$, $\sigma_{lnM}=3.84$ with $\chi^2 = 11.7$ and $\chi^2 = 43.2$ for $w_{qq}$ and $w_{qk}$ respectively. Finally, the open circles represent the $w_{qq}$ results scaled by a fitted factor of $\kappa/b=1/12=0.083\pm0.007$ for comparison with $w_{q\kappa}$ in the context of the WI98 model where galaxies are assumed to trace the mass.}
\label{fig:wqqwqkHOD_d12}
\end{figure}

\subsection{HOD parameter estimates for ATLAS QSOs}

The QSO-QSO $w_{qq}$ (blue) and the QSO-ACT $w_{q\kappa}$ (cyan) estimates are shown in Fig. \ref{fig:wqqwqkHOD_d12}. The $w_{qq}$ results are the same as those presented by \citetalias{Eltvedt_2}. Nevertheless, we first note that the HOD parameter fits of \citetalias{Eltvedt_2} (yellow line) now fail to fit these $w_{qq}$ data. This is because the CHOMP code used previously was not accurate at higher redshifts of the QSOs due to a comoving/proper distance `feature'. The HOD parameters of \cite{Petter2023} now also seem to underestimate $w_{qq}$ when we use them through CCL rather than CHOMP. However, since  Petter et al fitted a different QSO sample, this could explain the difference. But we shall see that the main effect of the change in our HOD fit is to move the peak halo effective mass to significantly higher ($\approx10\times$) masses  than we saw before.

We next  fit a HOD model to $w_{qq}$ here assuming that $logM_{min}=logM_0$ as assumed in \citetalias{Eltvedt_2} and we only fit in the range $\theta<20'$ excluding the two points at $\theta>20'$ on the grounds that they may be affected by systematics, given the low amplitude of $w_{qq}$ (see Fig. \ref{fig:wqqwqkHOD_d12}). We find a reasonable fit (red line), consistent with the idea that the 1-2 halo transition is at $\approx1'$ or $r\approx1$h$^{-1}$ Mpc (comoving) at $z\approx1.7$. The best fit HOD parameters for $w_{qq}$ are given in Table \ref{tab:hodfits}. 


We  also independently fit our observed QSO-ACT $w_{q\kappa}$, obtaining the HOD parameters also given in the caption of Fig. \ref{fig:wqqwqkHOD_d12} and Table \ref{tab:hodfits}. We  only fit in the range $\theta>3'$ excluding the points at smaller scales  on the grounds that they are affected by the ACT spatial resolution and our smoothing with a Gaussian filter of $FWHM=3'$ (see Sect. \ref{sec:ACTmap}). We find a reasonable fit (green line, $\chi^2=44.0$) whose parameters are given in Table \ref{tab:hodfits} for comparison with the $w_{qq}$ parameters and those for galaxies and LRGs.

To test for self-consistency between $w_{qq}$ and $w_{q\kappa}$ we then use the same $w_{qq}$ HOD parameters as estimated above  to predict $w_{q\kappa}$. After filtering in the range $40<l<3000$ we obtain the model shown in Fig. \ref{fig:wqqwqkHOD_d12} as the red dashed line. Considering points with $\theta>3'$ we see that the fit is reasonable out to $\theta<50'$, although a little too high at $\theta<50'$ and a little too low at $\theta>50'$. The overall $\chi^2=57.6$ for fitting to 17 points is only somewhat bigger than the  $\chi^2=44.0$ for the best $w_{q\kappa}$ fit and even closer given they have almost equal  reduced $\chi^2=3.38$ (=57.6/17 = 44.0/13). Finally, we further test the consistency of $w_{qq}$ and $w_{q\kappa}$ HOD parameter fits by fitting the combined $w_{qq}$ and $w_{q\kappa}$ data, finding a total $\chi^2=12.9+53.6=66.5$. This compares to $\chi^2=11.6+44.0=55.6$ when fitting $w_{q\kappa}$ directly and $\chi^2=11.6+57.6=69.2$ when fitting $w_{q\kappa}$ with the $w_{qq}$ fit. Overall, we conclude that it is easier to find self-consistent HOD parameter fits for the QSO auto- and cross-correlation functions than for the $17<r<21$ galaxies. In this case, the reason may be due to the correlation functions lying  almost wholly in the linear 2-halo regime and we have previously seen that Halofit models also gave reasonable small-scale fits to $w_{qq}$ and $w_{q\kappa}$. However, below we shall also test the alternative explanation that, like the LRGs, the QSO hosts are dominated by a single type of galaxy leading to better fits to single HOD models.

\begin{figure}
     \includegraphics[width=\columnwidth]{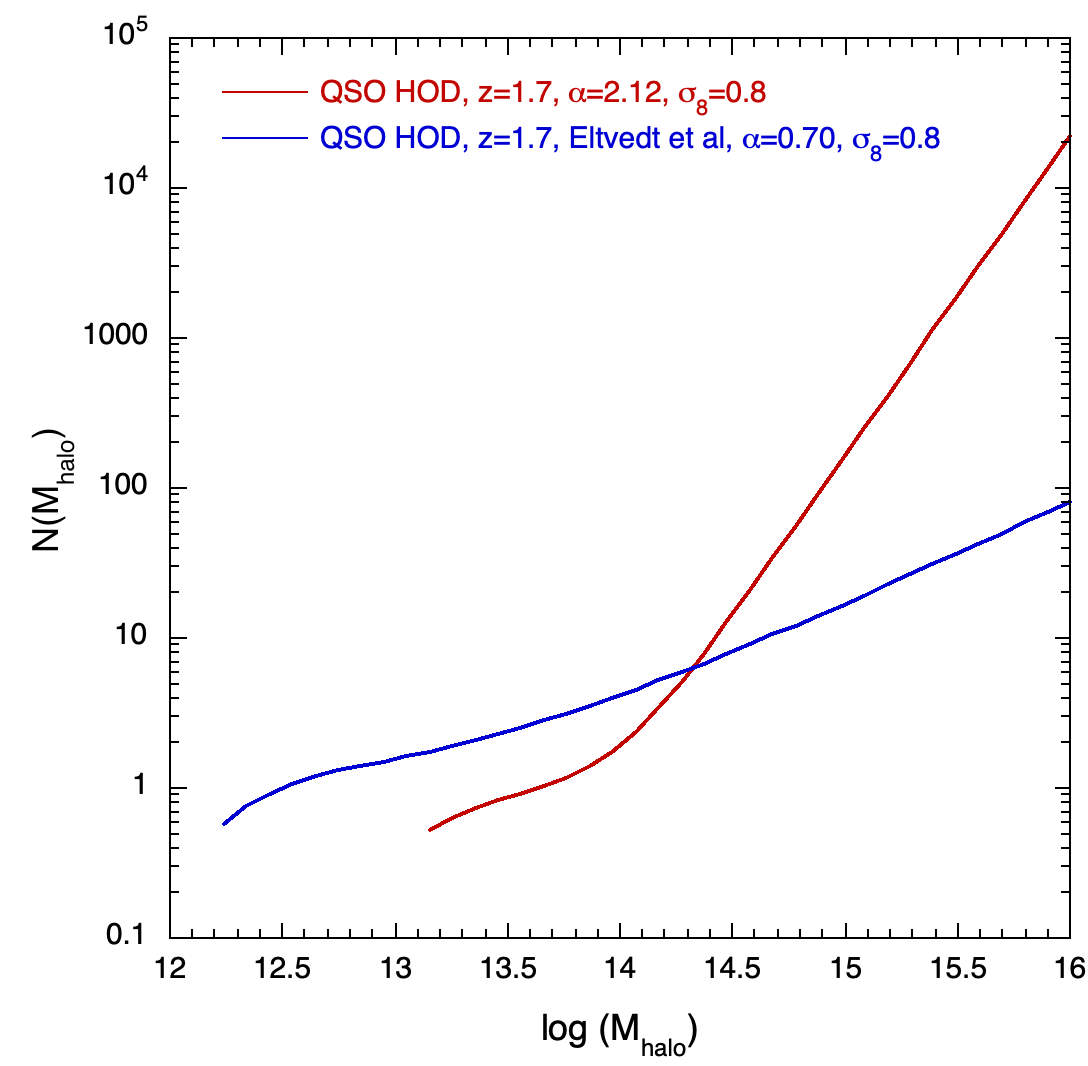}
     \includegraphics[width=\columnwidth]{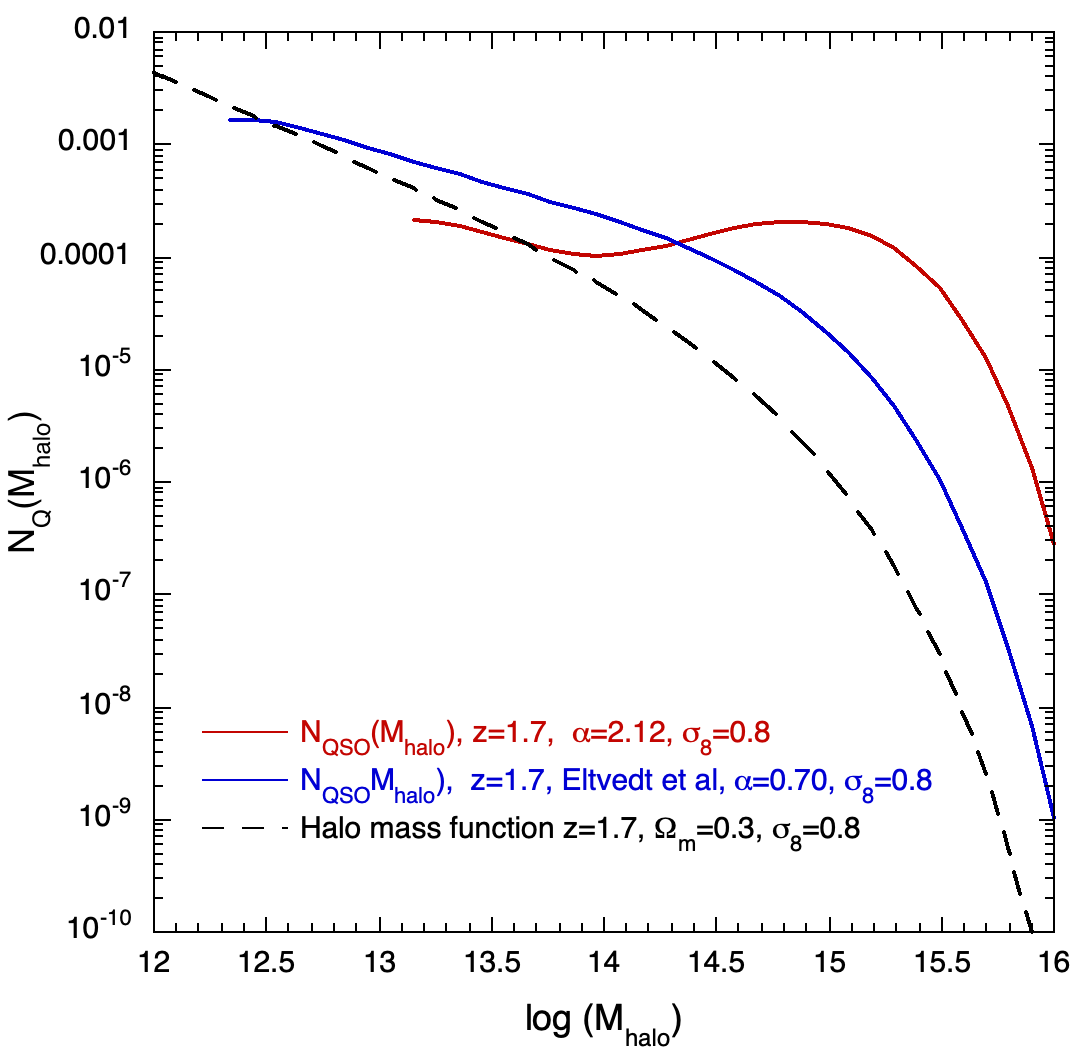}
     \caption{(a) HOD models are now shown for the $z\approx1.7$ QSOs based on the fitted parameters in Table \ref{tab:hodfits}. These are  $logM_{min}=13.13$, $logM_0=13.13$, ${logM_1}'=13.95$, $\alpha=0.2.12$, $\sigma_{lnM}=1.26$ and assuming $\sigma_8=0.8$ (red line). The blue line is the HOD model prediction from \protect\citetalias{Eltvedt_2} with $logM_{min}=12.2$, $logM_0=12.2$, ${logM_1}'=13.28$, $\alpha=0.70$, $\sigma_{lnM}=0.92$ and assuming $\sigma_8=0.8$. (b) The $z\approx1.7$ QSO $N_{QSO}$ - halo mass distributions, based on the above HODs represented in  (a) and (b)  by the red and blue lines, for comparison purposes.}
\label{fig:nq_mhalo_HOD_QSO}
\end{figure}

Fig. \ref{fig:nq_mhalo_HOD_QSO} (a) compares our QSO  HOD estimate (red line)with that of \citetalias{Eltvedt_3} (blue line) and we note that our current HOD estimate appears steeper than the previous HOD estimate. Fig. \ref{fig:nq_mhalo_HOD_QSO} (b) shows how this translates into the QSO host halo mass distributions after the halo mass function (dashed line) is multiplied by each HOD. We see that the QSO halo mass distribution now peaks at much  higher mass; the mean QSO host halo mass is now $\log M\approx 14.75$ compared to $\log(M)\approx12.4$ as quoted by \citetalias{Eltvedt_3}. This difference is surprisingly large and we emphasise that there is little constraint from the 1-halo term in our $w_{qq}$ fits and especially in our $w_{q\kappa}$ fits, despite the higher resolution of the ACT lensing map. Therefore we caution against the accuracy of this result. Nevertheless, this high effective mass from the HOD analysis is at least in the same direction as our $b_Q=4.44\pm0.24$ from our Halofit analysis which compares to $b_Q=2.09\pm0.3$ from comparing $w_{qq}$ to the expected  $\Lambda$CDM linear mass correlation function at $z=1.7$. As we noted in Section \ref{sec:halofit-qso}  and Fig.\ref{fig:sigma8_z} it is the lower $\sigma_8'(z=1.7)=0.23\pm0.03$ that we are measuring rather than assuming, that leads to our higher bias measurement than previously. But $b_Q=4.44$ still only corresponds to $13.4<logM<14.2$ for  the range $0.8<\sigma_8(z=0)<1.1$ in terms of the bias model of \cite{Tinker2010}. We conclude that the extra increase in the QSO $M_{eff}$ is due to the additional  effects included in the HOD modelling and not in the simple bias model. Thus QSOs appear to have an average host mass of $logM\approx14.75$ with $\approx2/3$ having $14<logM<16$ with $\approx1/3$ in the range $14<logM<14.75$ and $\approx1/3$ in the range $14.75<logM<16$.

\subsection{Non-linear bias estimates for ATLAS QSOs}

First, we tested a model where QSOs trace the mass, by scaling $w_{qq}$ downwards to best fit $w_{q\kappa}$. The blue open circles have therefore been scaled down by a factor of 12 and these seem to give a reasonable fit in the range $1<\theta<50'$. Thus $\kappa/b=1/12=0.083\pm0.007$ (fitting over the 7 points in the range $3'<\theta<30'$ for $\chi^2=4.04$)  and since $\kappa=0.33$ for QSOs at $1<z<2.0$,  then the ratio of the amplitudes of $w_{qq}$ and $w_{q\kappa}$ imply $b_Q=3.98$, again comparable to our previous result. Indeed, the continuing power-law excess at $\theta>50'$ in $w_{qq}$ appears to be reproduced in $w_{q\kappa}$, which may suggest that it is real, despite not being evident in the HOD (nor Halofit) models. Since, for the $17<r<21$ galaxies we found $\kappa=0.026$ and $\kappa/b=0.047$ giving $b_G=0.55$, assuming the amplitude of the observed $z\approx0.15$ acf's to be $r_0=5.0$ and $r_0=5.3$ for the QSOs we find a growth factor of $(5^{1.8}/0.55^2)/5.3^{1.8}/3.98^2)=59.9/1.27=47.2$, giving a growth rate of $D=\sqrt(47.2)=6.86$, high compared to an expected linear growth rate of 2.5 (or even 2.7 for EdS). However, at least at $z\approx0.15$, the $1<r<8$h$^{-1}$Mpc range is well into the non-linear regime where faster growth may be allowed and clearly seen in these observations.



Finally, we test a hypothesis that the QSOs may be hosted by LRGs.  Fig. \ref{fig:wqqwqkHOD_d12} shows a QSO model for $w_{qq}$ that assumes LRG HOD parameters at $z=1.7$ (solid cyan line) and similarly for $w_{q\kappa}$ at $z=1.7$. This model thus assumes no evolution in the HOD parameters in the range $0.26<z<1.7$ while allowing evolution in the underlying haloes based on the usual gravitational growth rate in an $\Lambda$CDM model. This model prediction for $w_{qq}$ is shown by the dashed mauve line in Fig. \ref{fig:wqqwqkHOD_d12} and the $w_{q\kappa}$ prediction by the mauve long-dashed line. After dividing the predicted $w_{qq}$ by a factor of 1.4  and $w_{q\kappa}$ by 1.2, we show these model predictions  in Fig. \ref{fig:wqqwqkHOD_d12} for $w_{qq}$ (mauve dashed line) $w_{qq}$ and $w_{q\kappa}$ (mauve long-dashed line). Although they get reasonably close to the observed $w_{qq}$ and $w_{q\kappa}$ estimates at scales $\theta>3'$, at smaller scales $w_{qq}$ shows an unobserved upturn due to the onset of the 1-halo regime. Unfortunately, the $w_{q\kappa}$ of the QSOs only reaches to $\theta>3'$ or $r>3$h$^{-1}$Mpc making it impossible to look for a similar up-turn in $w_{q\kappa}$. Meanwhile, on the basis of the small-scale $w_{qq}$ discrepancy, we conclude that QSO hosts may not be immediately identified as LRGs, although a higher resolution CMB lensing map is needed to further test this hypothesis.

Thus we conclude that QSOs can be self-consistently described by HOD models, although confined by our CMB map resolution to the linear 2-halo regime for $w_{q\kappa}$. We further conclude that a model where galaxies trace the mass is also a good fit to the $w_{qq}$ and $w_{q\kappa}$ data, although again higher resolution CMB lensing maps are needed to confirm this result.

\begin{figure}
     \includegraphics[width=\columnwidth]{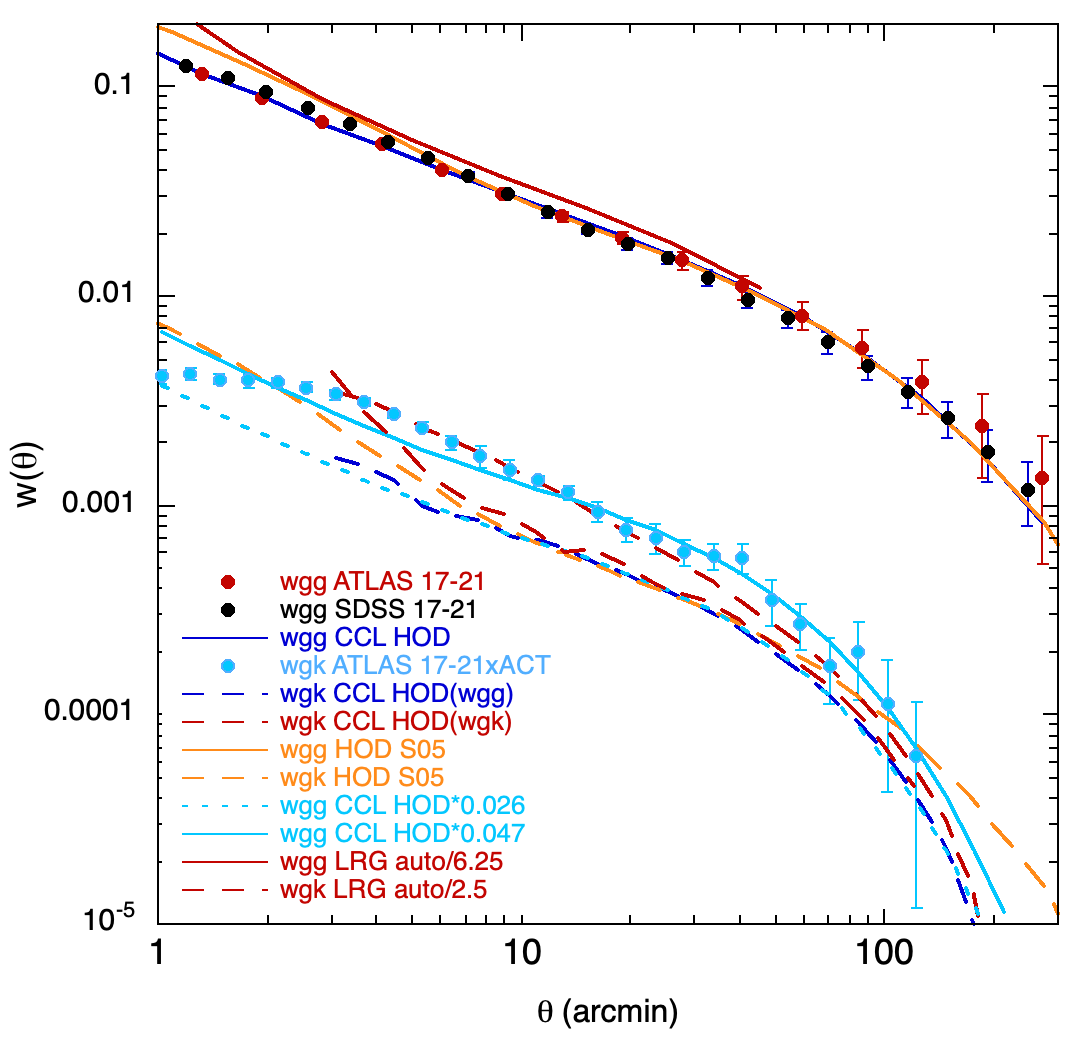}
     \caption{We compare  the 17<r<21 galaxy $w_{gg}$ and $w_{g\kappa}$ results from Fig. \ref{fig:wgg-wgk-hod} to the best fit LRG HOD model predictions multiplied by $0.4^2$ for $w_{gg}$ (red line) and by 0.4 for $w_{g\kappa}$ (red dashed line).  All other models have the same symbols as in Fig. \ref{fig:wgg-wgk-hod}. These `diluted LRG' HOD model predictions have the same problem as the  17<r<21 galaxy HOD models in their under-estimation of $w_{g\kappa}$ between $5'<\theta<40'$ by a factor of $\approx2$, suggesting that LRGs do not dominate the galaxy $w_{g\kappa}$ in this range.}
\label{fig:wgg_wgk_HOD_gal_LRG}
\end{figure}

\section{Discussion: 17<r<21 galaxies -  LRGs v. mass tracers}

\label{sec:discussion}

Similarly to what we have done above in terms of the QSOs, we now return to re-consider the $17<r<21$ galaxy results  to see if the galaxy population that dominates their  auto- and cross-correlation results might be LRG-like in the form of their mass profiles. We test this hypothesis by noting that we have to multiply the LRG HOD auto-correlation model in Fig. \ref{fig:wggwgkHOD_LRG} by $0.4^2$ to get an approximate fit to the $17<r<21$ galaxy $w_{gg}$ (see Fig. \ref{fig:wgg_wgk_HOD_gal_LRG}). Assuming a simple model where all mass is assumed to be  in LRG/early-type galaxies with none in later-types and that the LRGs/early-types form 40\% of  $17<r<21$ galaxies \citep{Metcalfe2001,Metcalfe2006} implies we have to multiply the LRG HOD prediction for $w_{g\kappa}$ by 0.4 for consistency and this is compared to the observed galaxy $w_{g\kappa}$ in Fig. \ref{fig:wgg_wgk_HOD_gal_LRG}. We immediately see that this `diluted' LRG HOD  model has the same problem as the HOD models we directly fitted to $w_{gg}$ and $w_{g\kappa}$ - if the HOD fits $w_{gg}$ then it underestimates $w_{g\kappa}$ by a factor of $\approx2$ in the $5'<\theta<40'$ or $1<r<8$h$^{-1}$Mpc range. Indeed, these two HOD models (galaxy and diluted LRG) are in good agreement with each other. We conclude that while LRGs could dominate the 17<r<21 galaxy HOD {\it model} prediction for $w_{g\kappa}$, they appear sub-dominant in terms of explaining its {\it observed} form.

Indeed, restricting our attention to the $5'<\theta<40'$ scales where anti-bias is most evident, we again note that the \cite{WI98} model with $b=0.55$ provides a much better fit to the $17<r<21$ galaxy $w_{gg}$ and $w_{g\kappa}$ than either the galaxy or LRG HOD models. Specifically, the galaxy-galaxy $\xi(r)$ having the same -1.8 slope as the galaxy-mass cross-correlation function suggests that this range is dominated by galaxies whose clustering traces the mass. This further implies that these galaxies have mass profiles that fall off more slowly than LRGs with NFW halo mass profiles; their asymptotic $\rho_(r)\propto r^{-3}$ would predict $w_{gg}\propto\theta^{-1}$ and $w_{g\kappa}\propto\theta^{-2}$, forms strongly rejected by our observed $w_{gg}\propto w_{g\kappa}\propto\theta^{-0.8}$. We suggest that a model similar to the hierarchical mass clustering model\footnote{Although the relation between the slope of $\xi(r) $ and the initial slope of the power spectrum would not apply in our assumed CDM model with initial adiabatic perturbations.} of \cite{Peebles1974}, rather than  a halo model, could solve this problem, since it is a model where galaxies trace the mass and predicts $w_{gg}\propto w_{g\kappa}\propto\theta^{-0.8}$, consistent with our observed results. Thus we conclude that while NFW mass profiles describe LRG mass profiles well, they may not provide  appropriate descriptions of the mass profiles and clustering of the galaxies that dominate in this $1<r<8$h$^{-1}$Mpc range. At these intermediate scales,  these other galaxies (a) appear to trace the mass and (b) have flatter power-law mass profiles with $\rho(r)\propto r^{-1.8}$. 

The main caveat is that the Halofit models in Fig. \ref{fig:wggwgk}, although designed to fit bias at large scales, also look like they are better fits than the HOD models at small scales. But between $1<\theta<8'$, the $w_{gg}$ model would have to be brought down from $b_A=1.05$ to $b_A=0.8$ to fit and the $b_X=1.0$ model bias (red line) would then have to be reduced by $\approx20$\% leaving it $\approx30-40$\% in total below the ACT $w_{g\kappa}$. Although this is less than the $\approx90$\% anti-bias found from the HOD or \cite{WI98} route, it has to be kept in mind that the Halofit model only parameterises the mass and not the galaxy distribution and so may be regarded as a model which intrinsically follows the mass. Thus it may be no surprise that it fits our $w_{gg}$ and $w_{g\kappa}$ results somewhere between the \cite{WI98} and HOD models.


\section{Conclusions}
\label{sec:conclusions}

We now summarise the main results from this paper where we have combined angular clustering analyses with angular ACT (and Planck)  CMB lensing cross-clustering analyses to VST ATLAS samples of $17<r<21$ galaxies, $0.16<z<0.36$ LRGs and $z\approx1.7$ QSOs. The conclusions are:-

1)  From our analysis of the angular auto- and cross-correlation functions of the $z\approx0.15$ galaxies, the $z\approx0.26$ LRGs and $z\approx1.7$ QSOs using a simple
Halofit plus bias model in the linear regime, we find bias values of respectively $b_G(z\approx0.15)=1.1$, $b_{LRG}(z=0.26)=2.76\pm0.11$ and $b_Q(z=1.7)=4.44\pm0.24$, using these we found $\sigma_8'(z=0.15)=0.66\pm0.053$, $\sigma_8'(z=0.26)=0.52\pm0.12$ and $\sigma_8'(z=1.7)=0.23\pm0.03$. We found from Fig. \ref{fig:sigma8_z} that $\sigma_8$ was increasing faster than expected in a $\lambda$CDM model and, indeed, more similar to the rate expected in an $\Omega_m=1$ Einstein de Sitter model. Since the lower redshift $z\approx0.15$ galaxy result for $\sigma_8(z=0)$ is reasonably consistent with  the $\sigma_8(z=0)=0.8$ value expected from $\Lambda$CDM, most of this effect arises from the lower than expected $\sigma_8(z=0)=0.49\pm0.06$ derived from the $z=1.7$ QSOs. While our results are lower than the QSO results of other authors, they are also in statistical agreement with those other results. 


2) We then fitted HOD models over a wider range of scales to these three samples. We note that the HOD parameters found here for the QSO sample disagree with those found in \citetalias{Eltvedt_3} for the same sample, due to a `feature' in the CHOMP code that did not treat high-z lens samples correctly. We found that within the \cite{Z2007} parameter HOD model formalism, the LRGs and QSOs allowed self-consistent fits for the auto- and cross-correlation functions to be derived. However, for  the lower redshift $z\approx0.15$ galaxy sample no self-consistent fit could be found with even the best fit to both $w_{gg}$ and $w_{q\kappa}$ leaving $w_{q\kappa}$ underestimated by a factor of $\approx2$ in the $1<\theta<40'$ range. We found a similar result when we fitted a model where galaxies traced the mass and found that we required $b=0.55$ to fit these data with the \cite{WI98} model.

3) The anti-bias of $b\approx0.5$ found for the $z\approx0.15$ sample is in excellent agreement  with  the results from the QSO lensing results of \cite{Myers2003,Myers2005} and \citetalias{Eltvedt_2} who also found $b\approx0.5$ in the range $1<\theta<5'$. \cite{Scranton2005} obtained very similar $w_{gq}$ results in their SDSS galaxy sample. They claimed that by using a simple HOD, they found self-consistent $w_{gg}$ and $w_{gq}$ results that agreed with standard $\Lambda$CDM model with no need for any anti-bias to obtain a fit but when we adjust their HOD to fit better their $17<r<21$ galaxy $w_{gg}$ result, their HOD model now also significantly underestimates $w_{gq}$ and also $w_{g\kappa}$.

4) We also tested whether the $z\approx0.15$ galaxy $w_{g\kappa}$ could be fitted by the HOD model that fitted the LRG sample $w_{g\kappa}$ but found that it could not. The preference for the \cite{WI98} model where galaxies trace the matter suggests that, rather than a halo model with an NFW mass profile the $z\approx0.15$ galaxies prefer a hierarchical clustering model of the type described by \cite{Peebles1974} where the density run around galaxies goes as $\rho(r)\propto r^{-1.8}$ and $w_{gg}\propto w_{g\kappa}\propto\theta^{-0.8}$.

5) Thus the $z\approx0.15$ galaxy sample was poorly fitted by HOD models and better fitted by models where galaxies traced the mass in the $1<r<8$h$^{-1}$Mpc range. The LRG sample was poorly fitted by the model where galaxies trace the mass and much better fitted by a HOD + NFW model. The QSO sample was reasonably fitted by a HOD model and also by the hierarchical  model of \cite{Peebles1974} where QSOs trace the mass under the \cite{WI98} formalism.

6) If an anti-bias of $b=0.55$ was found to be unphysical, then  there is the alternative of a significantly higher value for $\Omega_m=0.3/0.55\approx0.6$. This would be in tension with our own linear, Halofit estimates of $\Omega_m$ from the same data, as well as the  Hubble diagram estimates. However, we also find   support  for a higher $\Omega_m$ from the fast growth rate of $\sigma_8(z)$ in our linear clustering analyses, so we should remain open to this possibility, albeit at small odds.

7) Finally, higher resolution CMB data will allow important further tests of these conclusions. The 4MOST Cosmology survey will also provide accurate redshifts for our VST ATLAS galaxy, LRG and QSO samples allowing new z-space distortion tests of the competing explanations of  anti-bias versus $\Omega_m$ of the CMB lensing results. LSST and Euclid will allow further tests of the ACT and Planck CMB lensing map using the weak lensing shear technique in the same volumes as we have analysed here.


\section*{Acknowledgements}
Harshnoor Kaur thanks the Physics Dept. of Durham University for part funding of her MScR course.
We thank C.S. Frenk  and C.G. Lacey of Durham University for useful discussions.
We also thank Alice Eltvedt and Nigel Metcalfe (Durham University) for their help in preparing the VST ATLAS Survey and catalogues and B. Ansarinejad (Univ. of Melbourne) for initial help with the ACT/Planck data analysis. We acknowledge use of the ESO VLT Survey Telescope (VST) AT-
LAS. The ATLAS survey is based on data products from observations made with ESO Telescopes at the La Silla Paranal Observatory under
programme ID 177.A-3011(A, B, C, D, E, F, G, H, I, J, K, L, M, N; (see \citealt{shanks2015}.)
We acknowledge the use of data products from WISE, which is
a joint project of the University of California, Los Angeles, and the
Jet Propulsion Laboratory (JPL)/California Institute of Technology
(Caltech), funded by the National Aeronautics and Space Administration (NASA), and from NEOWISE, which is a JPL/Caltech project funded by NASA.
We acknowledge use of SDSS imaging and spectroscopic data.
Funding for SDSS-III has been provided by the Alfred P. Sloan
Foundation, the Participating Institutions, the National Science
Foundation, and the US Department of Energy Office of Science.
We finally acknowledge STFC consolidated grant ST/T000244/1
in supporting this research.
For the purpose of open access, the author has applied a Creative
Commons Attribution (CC BY) licence to any Author Accepted
Manuscript version arising.
\section*{Data Availability}

The ESO VST ATLAS and WISE data we have used are all publicly
available. The VST ATLAS QSO catalogue can be found at https:
//astro.dur.ac.uk/cea/vstatlas/qso catalogue/. All other data relevant
to this publication will be supplied on request to the authors.



\bibliographystyle{mnras}
\bibliography{bibliography_v3_TS} 

\bsp	
\label{lastpage}
\end{document}